\documentclass[usenatbib,useAMS]{mnras}
\usepackage{tablefootnote}
\usepackage{multirow}
\usepackage{xcolor}
\usepackage{hyperref}
\usepackage{graphicx}	
\usepackage{amsmath}	
\usepackage{amssymb}	
\usepackage{multicol}        
\usepackage{bm}		
\usepackage{url}
\usepackage[normalem]{ulem} 
\usepackage{caption}      
\usepackage{longtable}

\title[Radio properties of OH megamaser galaxies]{Radio continuum properties of OH megamaser galaxies}

\author[Yu. Sotnikova et al.]{
Yu.~V.~Sotnikova,$^1$\thanks{E-mail: lacertae999@gmail.com} 
Z.~Z.~Wu,$^{2}$
T.~V.~Mufakharov,$^{1,3,4}$
A.~G.~Mikhailov,$^{1}$
M.~G.~Mingaliev,$^{1,3}$
\newauthor
A.~K.~Erkenov,$^{1}$
T.~A.~Semenova,$^{1}$
N.~N.~Bursov,$^{1}$
R.~Y.~Udovitskiy,$^{1}$
V.~A.~Stolyarov,$^{1,3,5}$
\newauthor
P.~G.~Tsybulev,$^{1}$
Y.~J.~Chen,$^{4}$
J.~S.~Zhang,$^{6}$
Z.~Q.~Shen$^{4}$
and
D.~R.~Jiang$^{4}$
\\
$^{1}$Special Astrophysical Observatory of RAS, Nizhny Arkhyz 369167, Russia\\
$^{2}$College of Physics, Guizhou University, Guiyang 550025, China\\
$^{3}$Kazan Federal University, 18 Kremlyovskaya St, Kazan 420008, Russia\\
$^{4}$Shanghai Astronomical Observatory, Chinese Academy of Sciences, Shanghai 200030, China\\
$^{5}$Astrophysics Group, Cavendish Laboratory, University of Cambridge, J J Thomson Avenue, Cambridge CB3 0HE, UK \\
$^{6}$Center for Astrophysics, Guangzhou University, Guangzhou 510006, China \\     
}

\date{Received September 4, 2021; accepted xx, 2021}
\pubyear{2020}
\begin{document}
\label{firstpage}
\pagerange{\pageref{firstpage}--\pageref{lastpage}}
\maketitle

\begin{abstract}
We present a study of the radio continuum properties of two luminous/ultraluminous infrared galaxy samples: the OH megamaser (OHM) sample (74 objects) and the control sample (128 objects) without detected maser emission. We carried out pilot observations for 140 objects with the radio telescope RATAN-600 at 1.2, 2.3, 4.7, 8.2, 11.2, and 22.3 GHz in 2019--2021. The OHM sample has two times more flat-spectrum sources (32 per cent) than the control sample. Steep radio spectra prevail in both samples. The median spectral index at 4.7 GHz $\alpha_{4.7}=-0.59$ for the OHM sample, and $\alpha_{4.7}=-0.71$ for the non-OHM galaxies. We confirm a tight correlation of the far-infrared (FIR) and radio luminosities for the OHM sample. We found correlations between isotropic OH line luminosity $L_{OH}$ and the spectral index $\alpha_{4.7}$ ($\rho$=0.26, p-val.=0.04) and between $L_{OH}$ and radio luminosity $P_{1.4}$ ($\rho$=0.35, p-val.=0.005). Reviewing subsamples of masers powered by active galactic nuclei (AGNs) and star formation revealed insignificant differences for their FIR and radio properties. Nonetheless, AGN-powered galaxies exhibit larger scatter in a range of parameters and their standard deviations. The similarities in the radio and FIR properties in the two samples are presumably caused by the presence of a significant amount of AGN sources in both samples (47 and 30 per cent in the OHM and control samples) and/or possibly by the presence of undetected OH emission sources in the control sample.


\end{abstract}


\section{Introduction}
The galaxies hosting OH megamaser (OHM) emission are usually classified as the luminous or ultraluminous infrared galaxies (U/LIRGs), which contain a large amount of molecular gas in their central region \citep{2005ARA&A..43..625L,2019MNRAS.486.3350S}. Their high luminosities are explained by high rates of star formation or by the presence of active galactic nuclei (AGNs). Either way, these phenomena, resulting in outflows of matter into the interstellar medium or inflows towards the nuclear region, play a major role in galaxy evolution. \cite{2020MNRAS.498.2632H} present a hypothesis that OHM galaxies harbour recently triggered AGNs. Similarly, OHM galaxies could also be a transition stage between a starburst and the emergence of an AGN along the sequence of galaxy evolution \citep{2020A&A...638A..78P}. By investigating objects containing OHMs, it is possible to study the AGN’s influence on the processes of galaxy evolution that are triggered by the accretion of matter onto the supermassive black hole \citep{2015A&A...574A...4V}. 

One of the conditions to produce maser radiation in galaxies is the presence of background continuum radio emission that seeds the maser emission, since maser spectral lines are located in the radio frequency range \citep{2006A&A...449..559B}. Thus, measurements in the radio continuum are essential in determining the nature of the non-thermal radio emission and the mechanism for inducing maser radiation in OHMs. There is a strong correlation between the radio and infrared (IR) emission in star-forming (SF) galaxies (e.g. \citealt{1985ApJ...298L...7H,1991ApJ...376...95C,2003MNRAS.338..745Y}). The presence of luminous AGNs among IR galaxies can cause an excess of detected radio continuum compared to the level expected from star formation  \citep{2001ApJ...554..803Y}.
The radio spectral index is an important parameter that influences the correctness of calculated radio properties, because adoption of a fixed spectral index for all the sources leads to a significant scatter and a strong bias in the analysis as shown by \citet{2019ApJ...875...80G}.

\citet{2005Ap.....48...99K,2005Ap.....48..237K} 
studied the properties of 30 OHM galaxies in the radio continuum at a frequency of 1.49 GHz based on the VLA observations, it was the first work attempted to study a relatively large sample of OHMs with high angular resolution ($\sim1\arcsec$) in the radio continuum. The flat radio spectra and the high brightness temperatures obtained by the author suggest the presence of AGNs in these galaxies; however, the author could not rule out active star formation in some of the 30 studied OHMs. No significant correlation between the fluxes in the OH line and in the radio continuum was found for the studied sample. 
The author assumes that the thermal component of radio emission is insignificant at low frequencies and that synchrotron radiation predominates. Analysing the features of the megamasers in the list as well as the properties of the OH and far-infrared (FIR) luminosities correlation, the author suggests that saturated gain and collisional pumping are probably the preferred emission mechanism in OHMs.

The sample of OHM galaxies investigated by \citet{2005Ap.....48...99K,2005Ap.....48..237K} contains low-redshift (z $<$ 0.1) sources detected in 1980s. New discoveries available in the literature (e.g., \cite{2000AJ....119.3003D,2001AJ....121.1278D}) 
significantly expanded the number of known OHM galaxies. \cite{2014A&A...570A.110Z} compiled and studied two large samples of OHM and non-OHM galaxies and found differences in their middle-infrared (MIR) properties. The main objective of our study is a comparison of radio continuum properties for these OHM and non-OHM samples. 

We present pilot observations of OHMs made with the radio telescope RATAN-600 in 2019--2021. A catalogue of six-frequency flux densities and radio continuum spectra were obtained at 1.2, 2.3, 4.7, 8.2, 11.2, and 22.3 GHz quasi-simultaneously. For spectra classification, we compiled and analysed the average radio continuum spectra using both the RATAN observations and available literature radio data. We determined the radio properties of OMHs and compared them with the properties of a non-OHM control sample.

\section{The sample}

\cite{2014A&A...570A.110Z} compiled all 119 OH maser galaxies available in the literature (for 2014) and selected all the sources with no detection of maser emission among the updated Arecibo survey sample \citep{2002AJ....124..100D} as a control sample. We note that the selection criteria for that survey were $0.1 \leq z \leq 0.45$, $0\degr < \delta < 27\degr$, and InfraRed Astronomical Satellite (\textit{IRAS}) detection at 60~$\mu$m. We selected a total of 202 sources from \cite{2014A&A...570A.110Z} with 1.4 GHz flux densities $S_{1.4}> 5$ mJy for observations with RATAN: 74~OHM galaxies and 128 non-detections.

The OHM sample and control sample are presented in Table~\ref{tab:OHM} and Table~\ref{tab:control}, respectively, where Columns~(1)--(2) are the RA and Dec at 2000.0, Column~(3) is the redshift, Columns~(4)--(5) are the flux density at 1.4 GHz \citep{1998AJ....115.1693C} and the radio luminosity calculated at 1.4 GHz, Column~(6) is the logarithm of the radio-loudness, Columns~(7)--(8) are the spectral indices calculated at 1.4 and 4.7 GHz, and Column~(9) is the radio spectrum type.

According to the 13th edition of the catalogue of quasars and active nuclei by \cite{2010A&A...518A..10V}, there are 33 AGNs and 2 quasars among our OHMs (47 per cent of the sample), and 10 quasars, 27 AGNs, and one BL Lacertae-type object in the control sample (30 per cent of the sample).

\begin{figure}
\centerline{\includegraphics[width=.45\textwidth]{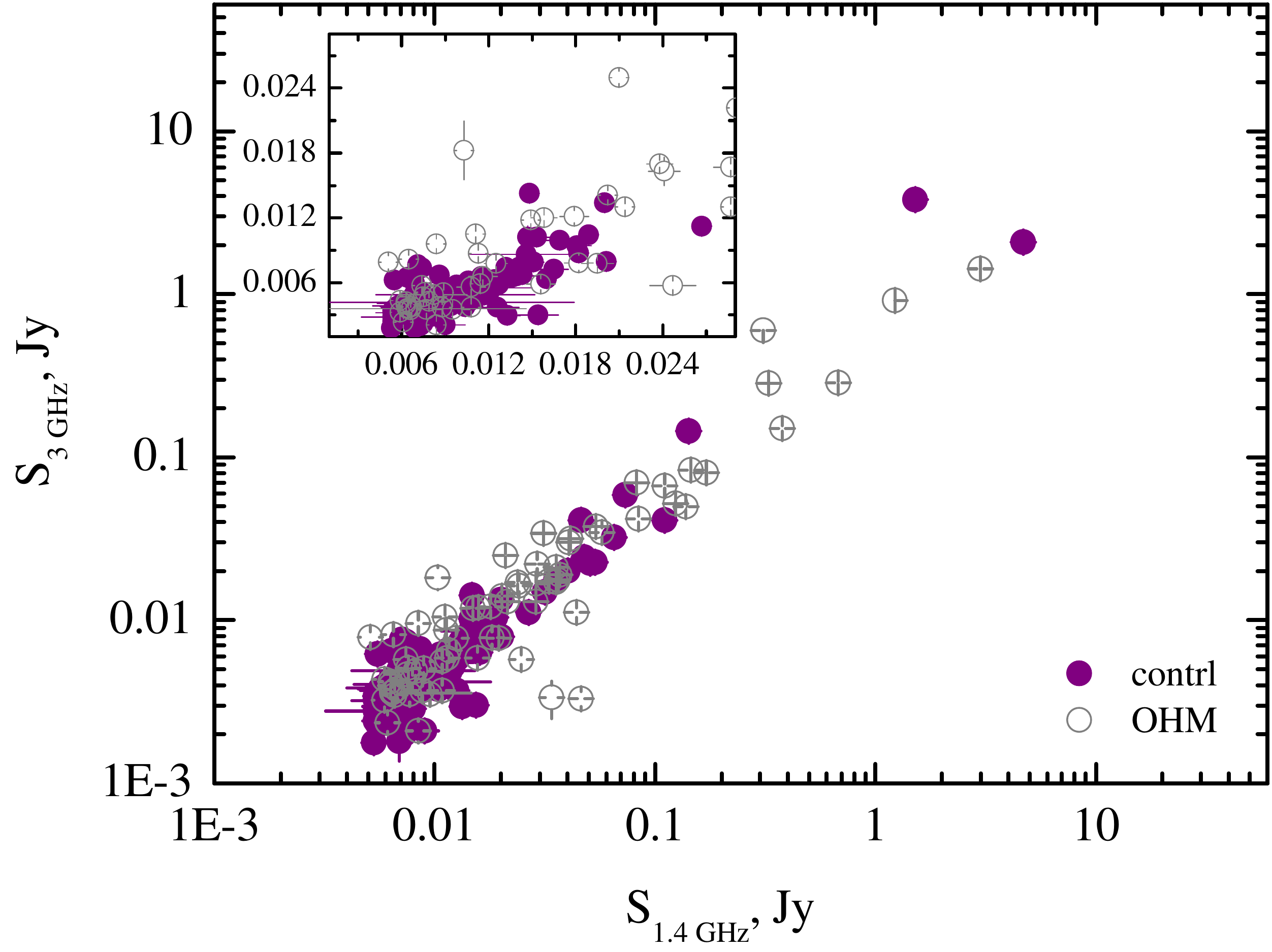}}
\caption{$S_{1.4}$ versus $S_{3}$ for the OHM and control samples.} 
\label{fig:flfl}
\end{figure}

The two samples under investigation are different in their flux density distribution. The flux--flux plot \mbox{$S_{1.4}$--$S_{3}$} (Fig.~\ref{fig:flfl}) is shown in the log scale due to a wide spread of objects' fluxes. Most of the sources have flux densities at 1.4 GHz less than 50 mJy (Fig.~\ref{fig:flfl}), the median value for the OHM sample is 19 mJy, for the control is 8.5 mJy. 

\begin{figure}
\centerline{\includegraphics[width=.44\textwidth]{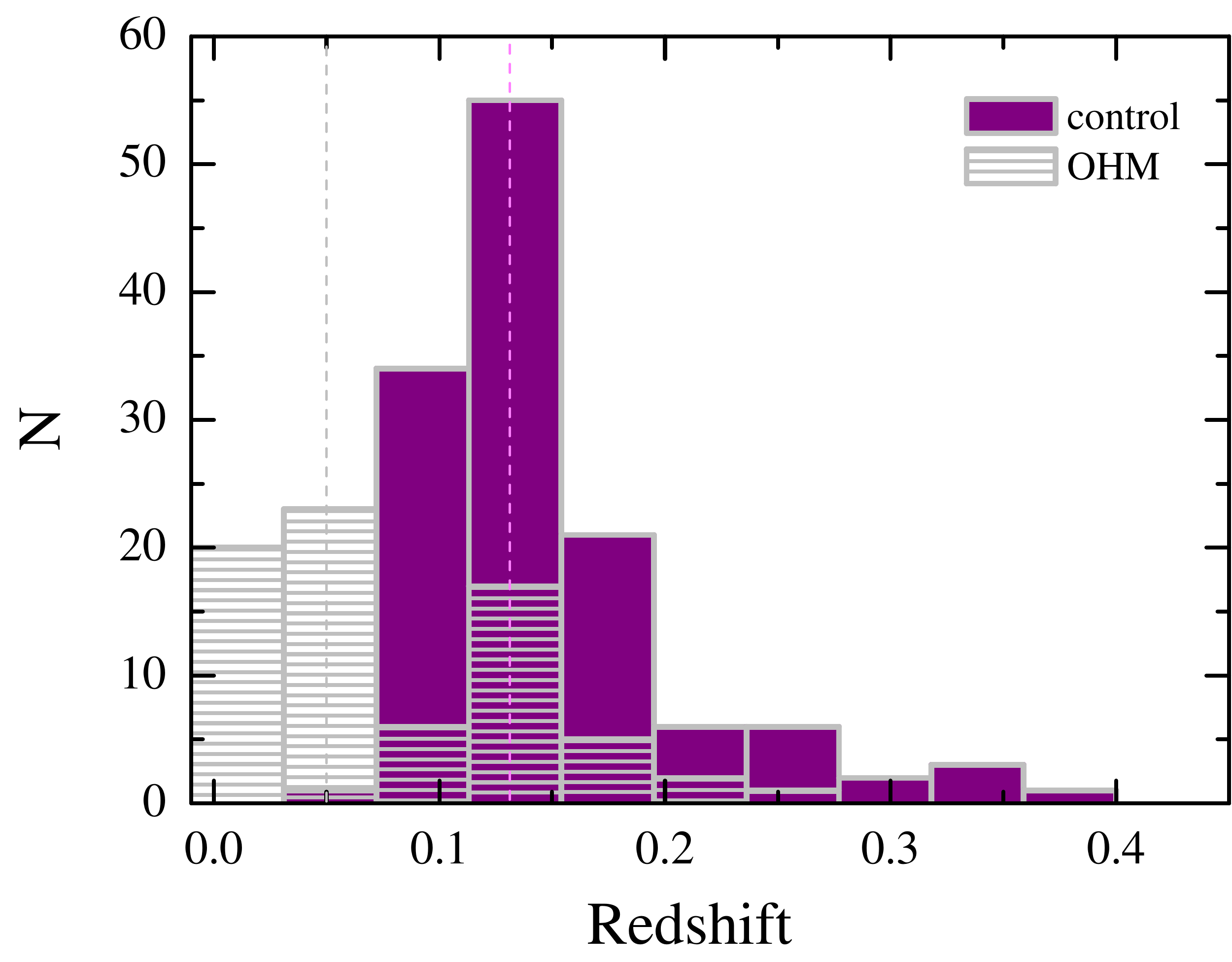}}
\caption{Redshift distributions for the OHM and control samples. The grey and pink lines are, respectively, the median redshifts for the samples.} 
\label{fig:z}
\end{figure}

Due to the weak flux densities, the sources from both samples have not been monitored systematically at radio frequencies, thus most objects have limited radio measurements. Almost all of them have two-point radio spectra at 1.4~GHz (NRAO VLA Sky Survey, NVSS; \citealt{1998AJ....115.1693C}) and 3 GHz (VLA Sky Survey, VLASS; \citealt{2020PASP..132c5001L}). The time interval between these measurements is 20 yr. The redshifts were taken from the National Aeronautics and Space Administration (NASA)/Infrared Processing and Analysis Center (IPAC) Extragalactic Database (NED),\!\footnote{https://ned.ipac.caltech.edu} and their median values are 0.05 for the OHMs sample and 0.13 for the control sample (Fig.~\ref{fig:z}).


\section{Observations and data reduction}

The radio telescope RATAN-600 \citep{1993IAPM...35....7P} with a \mbox{600-m} circular multi-element antenna operates in the transit mode and provides measurement of 1--22 GHz broad-band spectra. The multifrequency measurements of flux densities are obtained simultaneously within 1--2 min when a source moves along the focal line where receivers are located. The angular resolution (measured by the full width at half-maximum parameter, FWHM) in this mode depends on the antenna elevation angle, and the resolution along declination ${\rm FWHM}_{\rm Dec.}$ is three to five times worse than that along right ascension ${\rm FWHM}_{\rm RA}$. The angular resolution along RA and Dec. calculated for the average angles is presented in Table~\ref{tab:radiometers} for different frequencies. The detection limit for a RATAN-600 single sector is approximately 8 mJy at 4.7 GHz (integration time is about 3 s) under good conditions and at the average antenna elevation angle (${\rm Dec.} \sim 0\degr$). It also depends on the atmospheric extinction and the effective area for different antenna elevation angles $H$ (from 10$\degr$ up to 90$\degr$) at corresponding frequencies.

The observations for the sample of galaxies were obtained in 2019--2021. We observed the sources from 3 to 40~times for each observing epoch to improve the signal-to-noise (S/N) ratio. The observations were carried out with the radiometric array at six frequencies: 1.25, 2.25, 4.7, 8.2, 11.2, and 22.3 GHz. The measurements were processed using the automated data reduction system \citep{2011AstBu..66..109T,2016AstBu..71..496U,2018AstBu..73..494T} and the Flexible Astronomical Data Processing System (\textsc{FADPS}) standard package modules \citep{1997ASPC..125...46V} for the broad-band RATAN continuum radiometers. We used the following eight flux density secondary calibrators: 3C48, 3C138, 3C147, 3C161, 3C286, and NGC\,7027. The flux density scales calculated based on \cite{1977A&A....61...99B} and \cite{2013ApJS..204...19P,2017ApJS..230....7P} are in good agreement with each other, and their differences are small compared to the measurement errors. Additionally, we used the traditional RATAN-600 flux density calibrators: J0240$-$23 and J1154$-$35 \citep{2019AstBu..74..497S}. The measurements of the calibrators were corrected for the angular size and linear polarization according to the data from \citet{1994A&A...284..331O} and \citet{1980A&AS...39..379T}.

\begin{figure}
\centerline{\includegraphics[width=.44\textwidth]{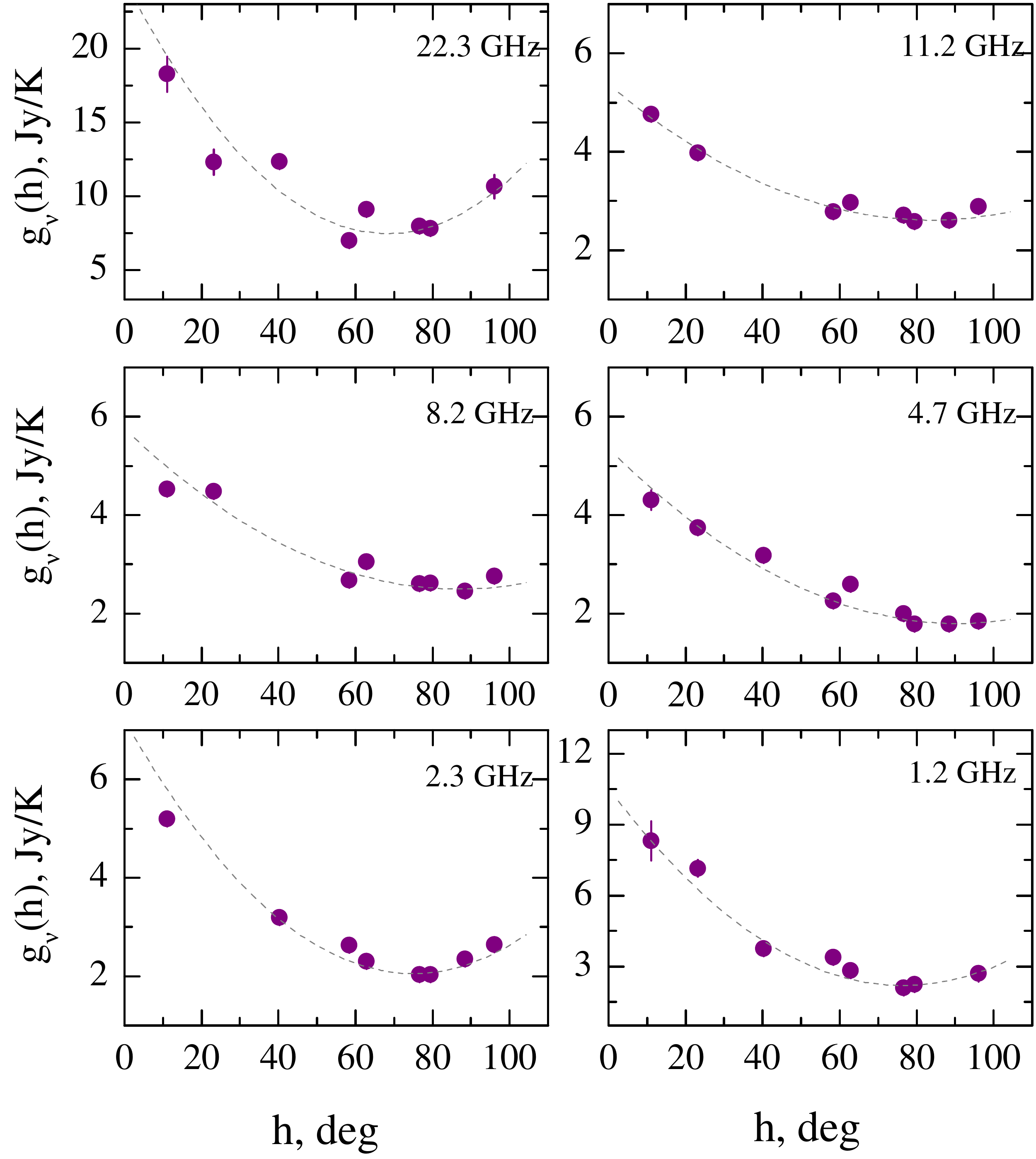}}
\caption{Flux density calibration factor $g_{\nu}$ versus antenna elevation angle $h$ at our six frequencies. All data for 8 calibrators were averaged over one observing run in 2020 July.}
\label{fig:calibrations}
\end{figure}

The total flux density error, which includes the uncertainty of the RATAN-600 calibration curve and the error in the antenna temperature measurement \citep{2012A&A...544A..25M,2016AstBu..71..496U}, is calculated by the following equation:
\begin{equation}
\left(\frac{\sigma_{S}}{S_{\nu}}\right)^2 = \left(\frac{\sigma_{c}}{g_{\nu}(h)}\right)^2 + \left(\frac{\sigma_{m}}{T_{{\rm ant},\nu}}\right)^2,
\end{equation}
where $\sigma_{S}$ is the total standard flux-density error;
$S_{\nu}$, the flux density at a frequency ${\nu}$;
$\sigma_{c}$, the standard calibration curve error, which is about 1--2 per cent and 2--5 
per cent at 4.7 and 8.2 GHz, respectively;
$g_{\nu}(h)$, the elevation angle calibration function (Figure~\ref{fig:calibrations});
$\sigma_{m}$, the standard error of the antenna temperature $T_{\rm ant}$ measurement;
and $T_{{\rm ant},\nu}$ is an antenna temperature.

The uncertainty of the antenna temperature measurement depends on the receiver's noise, the atmospheric fluctuations, and the accuracy of the antenna surface setting for an actual source observation. The systematic uncertainty of the absolute flux density scale (3--10 per cent
at 1--22 GHz) is not included in the total flux error.

\begin{table}
\caption{RATAN-600 continuum radiometer parameters: the central frequency $f_0$, the bandwidth $\Delta f_0$, the detection limit for point sources per transit $\Delta F$. ${\rm FWHM}_{\rm {RA} \times \rm {Dec.}}$ is the angular resolution along RA and Dec. calculated for the average angles.} 
\label{tab:radiometers}
\centering
\begin{tabular}{cccr@{$\,\times\,$}l}
\hline
$f_{0}$ & $\Delta f_{0}$ & $\Delta F$ &  \multicolumn{2}{c}{FWHM$_{\rm {RA} \times \rm{Dec.}}$}\\
GHz    &   GHz           &  mJy/beam   &   \multicolumn{2}{c}{}  \\
\hline
 $22.3$ & $2.5$  &  $50$ & $0\farcm17$ & $1\farcm6$  \\ 
 $11.2$ & $1.4$  &  $15$ & $0\farcm34$ & $3\farcm2$ \\ 
 $8.2$  & $1.0$  &  $10$ & $0\farcm47$ & $4\farcm4$   \\ 
 $4.7$  & $0.6$  &  $8$  & $0\farcm81$ & $7\farcm6$   \\ 
 $2.25$  & $0.08$  &  $40$ & $1\farcm71$ & $15\farcm8$  \\ 
 $1.25$  & $0.08$ &  $200$ & $3\farcm07$ & $27\farcm2$ \\ 
\hline
\end{tabular}
\end{table} 

\section{Radio data from the literature}

The literature radio data (Table~\ref{tab:catalogs}) are very rare, and there are only two catalogues for all the sources under investigation. They are the NVSS at 1.4~GHz in 1996 \citep{1998AJ....115.1693C} and the VLASS\footnote{https://cirada.ca/catalogues} at 3~GHz in 2017 \citep{2020PASP..132c5001L,2020RNAAS...4..175G}. Using the Astrophysical CATalogs support System (CATS) data base \citep{1997BaltA...6..275V, 2005BSAO...58..118V}, we complemented the compiled radio spectra in the samples with the following additional radio data: the Green Bank 6-cm survey (GB6, \citealt{1996ApJS..103..427G}), the GaLactic and Extragalactic All-sky Murchison Widefield Array (GLEAM) survey at \mbox{72--231}~MHz (2013--2014, \citealt{2017MNRAS.464.1146H}), the Giant Metrewave Radio Telescope Sky Survey (TGSS) at 150 MHz (2015, \citealt{2017A&A...598A..78I}), and the VLA measurements \citep{2007ApJS..171...61H}.

To find the counterparts of target objects in a large number of the radio catalogues, the {\it astroquery} Python package was used \citep{2019AJ....157...98G}. Radio fluxes for the OHMs and control sample objects were extracted from NED and VizieR Information System\footnote{https://vizier.u-strasbg.fr/viz-bin/VizieR} \citep{2000A&AS..143...23O}. The most useful sources of data in VizieR for OHMs turned out to be the papers of \cite{1990ApJS...73..359C}, \cite{1991ApJ...378...65C}, \cite{2002AJ....124..100D}, \cite{2003A&A...409..115N}, \cite{2015A&A...574A...4V}, and \cite{2018MNRAS.474.5008D}. 

\begin{table}
\caption{\label{tab:catalogs}The main external radio catalogues}
\centering
\begin{tabular}{ccl}
\hline
\multirow{2}{*}{Survey} & Frequency &  \multirow{2}{*}{Reference} \\
       &    (GHz)    &            \\
\hline
NVSS  & 1.4       & \cite{1998AJ....115.1693C} \\  
GLEAM & 0.072--0.231 & \cite{2017MNRAS.464.1146H} \\
TGSS  & 0.15       & {\cite{2017A&A...598A..78I}} \\
VLASS & 2--4        & \cite{2020PASP..132c5001L} \\
GB6   & 4.8        & \cite{1996ApJS..103..427G} \\ 
VLA   & 8.4        & \cite{2007ApJS..171...61H} \\ 
\hline
\end{tabular}
\end{table}

\section{Results}

\subsection{Flux densities}
\label{flux}
Table~\ref{tab:OHMfluxes} and Table~\ref{tab:controlfluxes} (fragments)
present the flux densities obtained with RATAN-600: Column 1 is the NVSS source name, Column 2 is the averaged observing epoch (JD), Column 3 is the number of observations $N_{obs}$, Columns 4--9 are the flux densities and their uncertainties (Jy) at the RATAN-600 six frequencies. The dashes indicate non-detection cases, the symbol ``$<$'' denotes an upper flux density limit at 4.7 GHz. The data are presented in the VizieR Information System. In 2019--2021 we observed 83 objects from the control sample (65 per cent) and 57 ones (77 per cent) from the OHM sample. We have found that the detection rate at the RATAN frequencies is low. For the two samples it is 6, 29, 37, 76, 14, and 4 per cent at 22.3, 11.2, 8.2, 4.7, 2.3, and 1.2~GHz, respectively. The detection rates for each sample are presented in Table~\ref{tab:detection}. In general, the sources in both samples are quite faint radio sources with flux densities less that 10 mJy at radio frequencies.

In many cases, the absence of data at certain frequencies is a result of exclusion due to sensitivity and/or a strong influence of the man-made interferences at 1.2, 2.3, and 11.2 GHz ($-10\degr \leq \rm Dec. \leq 0\degr$). The flux density standard errors are \mbox{7--26}~per cent for the control sample and 10--25 per cent for the OHM sample. Table~\ref{tab:detection} presents the measurement uncertainties for the six frequencies.

\begin{figure}
\centerline{\includegraphics[width=.37\textwidth]{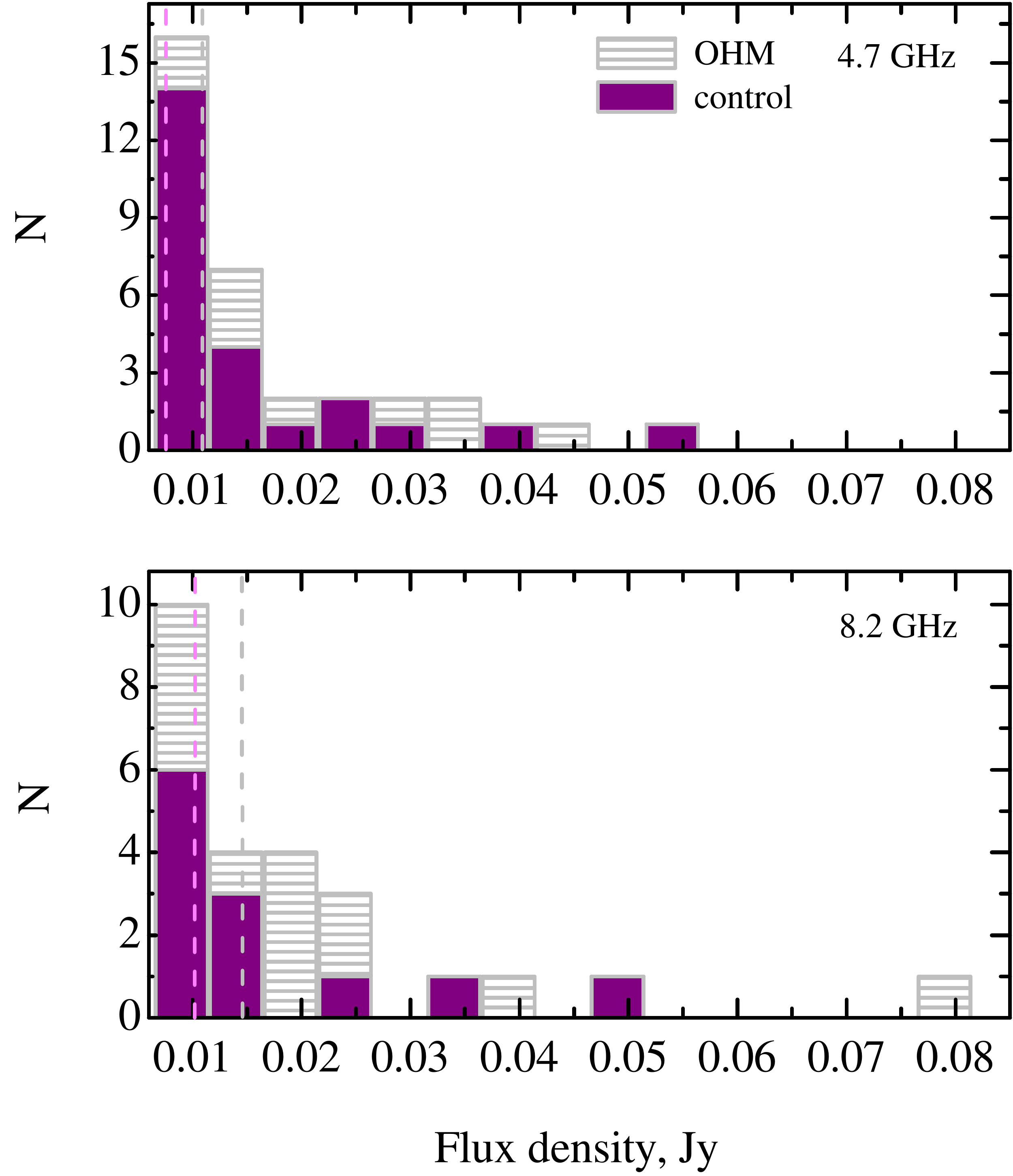}}
\caption{The OHM flux density distributions at 4.7 (top) and 8.2 GHz (bottom) for the control (purple) and maser (grey) samples; the pink and grey lines are, respectively, the median flux densities for the samples. 10 brightest radio sources from both samples, J0134+32, J0851+20, J1226+02, J1345+12, J2314+03, J0045$-$25, J0240$-$00, J0248+43, J0331$-$36, and J1532+23, were excluded from the histogram.}
\label{fig:fluxes}
\end{figure}

The samples contain several brightest radio sources that influence their average fluxes. There are five sources in the control sample with $S_{\rm radio} > 1$ Jy (J0134+32, J0851+20, J1226+02, J1345+12, and J2314+03) and two very bright sources (J0045$-$25, J0240$-$00) in the OHM sample. The flux density statistics for the two samples is presented in Table~\ref{tab:detection}. In Fig.~\ref{fig:fluxes} the flux density distributions at 4.7 and 8.2 GHz for the samples are shown. 
 
We calculated the median flux densities at 8.2 GHz to be 0.015 and 0.011 Jy for the OHM and control samples. At 4.7 GHz they are 0.011 and 0.007 Jy, respectively. The Kolmogorov--Smirnov test shows that the flux density distributions at 4.7, 8.2, and 11.2 GHz for the OHM and control samples are significantly different at the level of 0.05. 

\begin{table*}
\caption{\label{tab:OHMfluxes} RATAN-600 flux densities for the OHM galaxies, measured simultaneously at six frequencies (1.2--22.3 GHz). Column 1 is the NVSS source name, Column 2 is the averaged observing epoch (JD), Column 3 is the number of observations $N_{obs}$, Columns 4--9 are the flux densities and their uncertainties (Jy). 
The dashes indicate non-detection cases. For the sources for which there is no detection at any RATAN frequency, we provide upper flux density limits at 4.7 GHz denoted by the symbol ``$<$''. The full version is available as supplementary material.}
\begin{tabular}{lcrrrrrrr}
\hline
Name & JD & $N_{obs}$ & $S_{22.3}$ & $S_{11.2}$ & $S_{8.2}$ & $S_{4.7}$ & $S_{2.3}$ & $S_{1.2}$ \\
\hline
J003600-271535 & 2458878 & 9  &   --             &      --          &  --    & 0.007$\pm$0.002 &  --              &  --              \\
J004733-251717 & 2459060 & 5  &   0.560$\pm$0.100 & 0.829$\pm$0.100   & 1.125$\pm$0.100   & 1.361$\pm$0.100 & 1.770$\pm$0.100  &  --              \\
J005334+124133   & 2459073 & 10 &   --             &      --          & --              & 0.004$\pm$0.002 &  --              &  --              \\
J014430+170607   & 2459071 & 6  &   --          & 0.012$\pm$0.004 & 0.022$\pm$0.005     & 0.026$\pm$0.004 &  --              &  --              \\
J024240-000047 & 2459047 & 8  &   0.412$\pm$0.100 & 0.644$\pm$0.100   & 0.944$\pm$0.100   & 1.527$\pm$0.100 & 3.150$\pm$0.200  &  3.160$\pm$0.300 \\
...\\
J132732+473906	& 2459007  & 7  & -- & -- &	-- & $<$0.006	& -- & -- \\
\hline
\end{tabular}
\end{table*}

\begin{table*}
\caption{\label{tab:controlfluxes} RATAN-600 flux densities for 
the control sample galaxies, measured simultaneously at six frequencies (1.2--22 GHz). Column 1 is the NVSS source name, Column 2 is the averaged observing epoch (JD), Column 3 is the number of observations $N_{obs}$, Columns 4--9 are the flux densities and their uncertainties (Jy). The dashes indicate non-detection cases. For the sources for which there is no detection at any RATAN frequency, we provide upper flux density limits at 4.7 GHz denoted by the symbol ``$<$''. The full version is available as supplementary material.}
\begin{tabular}{lcrrrrrrr}
\hline
Name & JD & $N_{obs}$ & $S_{22.3}$ & $S_{11.2}$ & $S_{8.2}$ & $S_{4.7}$ & $S_{2.3}$ & $S_{1.2}$ \\
\hline
J000438+365330 & 2459238 & 29 &  -- & --  & -- & 0.004$\pm$0.002 & --  & --  \\
J002930+243008 & 2459068 & 24 &  -- & --  & -- & 0.005$\pm$0.002 & --  & --  \\
J015328+260939 & 2459230 & 9 & -- & 0.009$\pm$0.003& --& 0.015$\pm$0.003& --& --\\
J015950+002338 & 2459249 & 17 & --  & 0.006$\pm$0.003& 0.007$\pm$0.002 & 0.009$\pm$0.002 & -- & -- \\
... \\
J024346+040636 & 2459068 & 11 & -- & -- & -- & $<$0.005 & -- & -- \\
\hline
\end{tabular}
\end{table*}

\begin{table}
\caption{\label{tab:detection} Statistics of the RATAN-600 measured flux densities. The columns are the following: (1) the observed frequency $f$ (GHz), (2) the detection rate, (3) the mean flux density $\tilde{S}$ (Jy) and its standard deviation (in parentheses), (4) the median flux density (Jy).}
\centering
\begin{tabular}{ccrc}
\hline
$f$    & Detection rate &  $\tilde{S}$ ($\sigma_{S}$) & Median \\
(GHz)  &    (per cent)        &  (Jy)                       & (Jy)   \\ 
\hline                                
\multicolumn{4}{c}{control} \\
\hline
22.3      & 6 & 4.168 (5.34)  &  1.223 \\ 
11.2      & 23 & 1.365 (3.85) &  0.011 \\ 
8.2       & 27 & 1.340 (4.28) &  0.011 \\ 
4.7       & 67 & 0.718 (3.56) &  0.007 \\ 
2.3       & 12 & 5.872 (11.81) & 1.300 \\ 
\hline
\multicolumn{4}{c}{OHM}  \\
\hline
22.3      & 7  & 0.389 (0.24) & 0.474 \\ 
11.2      & 39 & 0.113 (0.24) & 0.014 \\ 
8.2       & 51 & 0.115 (0.29) & 0.015 \\ 
4.7       & 89 & 0.090 (0.30) & 0.011 \\ 
2.3       & 18 & 0.639 (1.05) & 0.011 \\ 
1.2       & 2  & 3.155 --  &  3.155 \\ 
\hline
\end{tabular}
\end{table}

\subsection{Radio spectra}
\label{radiospectra}

The broad-band radio spectra of the OHMs and the galaxies from the control sample are shown in Figures \ref{fig:spectra1}-\ref{fig:spectra10}. They are compiled using the quasi-simultaneous RATAN measurements in 2019--2021 (shown in red colour) and non-simultaneous radio data from the literature (shown in black colour), which cover a time period of 1996--2020.
The low-frequency data are well represented by the GLEAM survey at 72--231 MHz (2013--2014; \citealt{2017MNRAS.464.1146H}) and TGSS at 150 MHz (2015; \citealt{2017A&A...598A..78I}).

\begin{figure}
\centerline{\includegraphics[width=.30\textwidth]{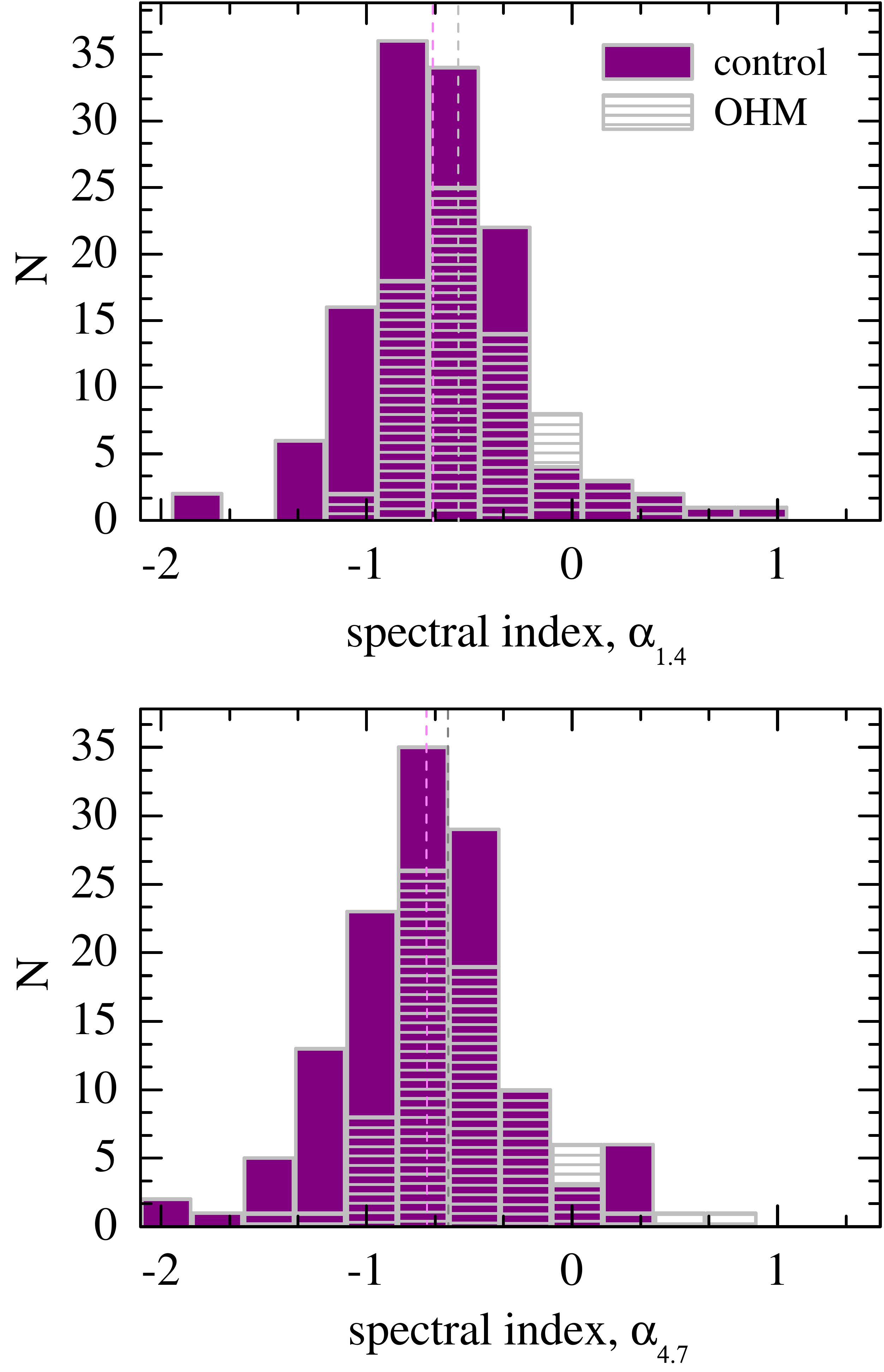}}
\caption{Distributions of the 1.4 and 4.7 GHz spectral indices for the OHM and control samples. The grey and pink lines are the median spectral indices, which are $\alpha_{1.4}=-0.55$ and $-$0.70, and $\alpha_{4.7}=-0.59$ and $-$0.71 for the OHM and control sample, respectively.}
\label{fig:indices}
\end{figure}

\begin{figure}
\centerline{\includegraphics[width=.45\textwidth]{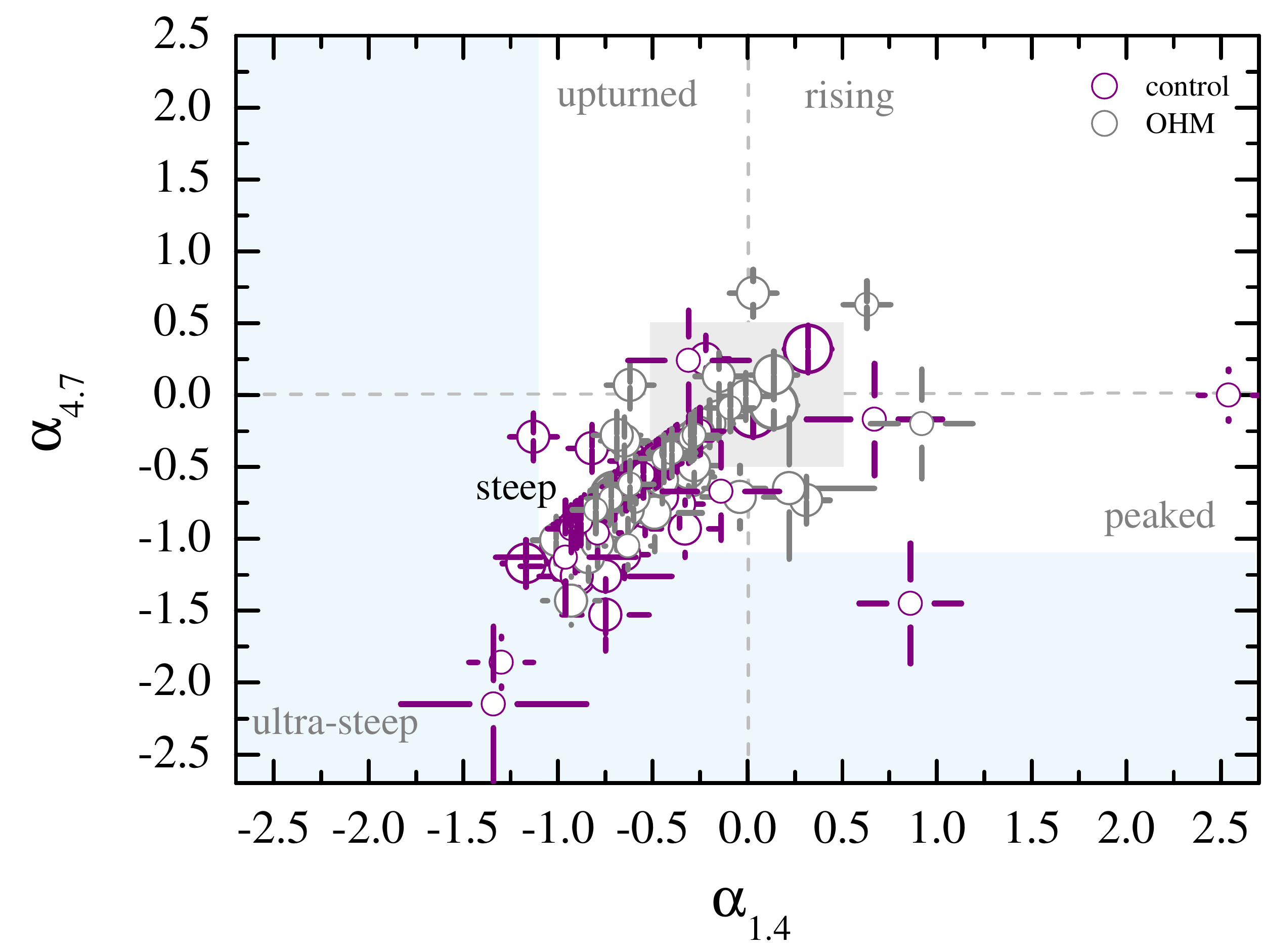}}
\caption{Radio two-colour diagram for the OHM (grey) and control (purple) samples. The blazar box is shown by the grey colour, the ultra-steep spectra area is presented by the blue colour. Objects with data points at only two frequencies are excluded from the diagram.}
\label{fig:two-color}
\end{figure}

We analysed the averaged radio spectra of the objects by fitting polynomials or straight lines (the blue lines on the spectra), calculated by weight averaging the measurements in the \textsc{FADPS} software package \citep{1997BaltA...6..275V}, and determined the spectral indices $\alpha$ in the power law $S_{\nu} \propto \nu^{\alpha}$, where $S_{\nu}$ is the flux density at a frequency $\nu$. In Figure~\ref{fig:indices} we present the distributions of the spectral indices at 1.4 and 4.7~GHz. The spectral types for the objects from both samples are presented in a radio two-colour diagram in Figure~\ref{fig:two-color}, where $\alpha_{1.4}$ is plotted against $\alpha_{4.7}$. The symbol sizes are proportional to the radio flux density at 1.4 GHz. 10 OHMs and 56 non-OHMs have radio data at only two frequencies and they were excluded from this 
diagram to avoid uncertain spectral classification due to variability at a large time-scale (about 20 yr). For both samples the $\alpha_{1.4}$ and $\alpha_{4.7}$ distributions are not significantly different at the 0.05 level 
according to the Kolmogorov--Smirnov test before and after excluding the sources with two-point spectra.

Table~\ref{tab:type} presents the spectral types in the OHM and control samples and the accepted classification criteria. We determined the spectral indices $\alpha_{\rm low}$ and $\alpha_{\rm high}$ for the peaked and upturn spectra by the frequencies lower or higher the point where the spectral slope changes its sign from positive to negative or vice versa. For the rest of the spectra, we calculated the low-frequency spectral index between the TGSS, GLEAM, WENSS, and other decimetre-wavelength measurements, while the high-frequency spectral index was estimated using the centimetre wavelength range. We considered a spectrum to be flat for spectral indices $-0.5 \leq \alpha \leq 0$, and rising if $\alpha > 0$. A spectrum was considered steep or ultra-steep for $-1.1 < \alpha < -0.5$ and $\alpha \leq -1.1$, respectively (see Table~\ref{tab:type}).

In Table~\ref{tab:stats}, we present the mean and median spectral index values for the two samples. Both samples are represented mostly by objects with steep spectra ($\alpha < -0.5$) at 1.4 and 4.7 GHz, but the
OHMs have a significant number of flat spectra (32 per cent). Also, the distribution of the spectral indices for the OHMs
is more compact with less standard deviation and less scatter between the maximum and minimum values. In both samples, there is no significant difference between the spectral indices at 1.4~GHz and at 4.7 GHz. It is caused by a large number (about 33 per cent) of spectra where spectral fitting was made with the single linear component due to the lack of radio measurements beyond the considered frequencies. We have found that the distributions of both the 1.4 and 4.7~GHz spectral indices are different for the OHM sample and for the control sample at a significance level of 0.05 according to the Kolmogorov--Smirnov test.

 \begin{table}
 \begin{small}
 \caption{\label{tab:type}Spectral types in the control and OHM samples.}
 \centering
 \begin{tabular}{lcrr}
 \hline
 Type & Criteria  &  $N_{\rm contr}$, \% & $N_{\rm OHM}$, \% \\
 \hline
 Peaked      & $\alpha_{\rm low}>0$, $\alpha_{\rm high}<0$ & 2.3 & 5.4\\ 
 Flat        & $-0.5 \leq \alpha \leq 0$                 & 17.2  &  32.4\\ 
 Rising    & $\alpha_{\rm low}>0$, $\alpha_{\rm high}>0$ & 5.5 & 5.4 \\ 
 Upturn      & $\alpha_{\rm low}<0$, $\alpha_{\rm high}>0$ & 1.6 & 2.7  \\
 Steep       & $-1.1 < \alpha < -0.5$ & 61.7 &  52.7                     \\ 
 Ultra-steep & $\alpha \leq -1.1$ &  11.7  &  1.4                    \\ 
\hline
\end{tabular}
\end{small}
\end{table}

\begin{table*}
\caption{\label{tab:stats} Statistics of spectral indices, radio loudness, FIR luminosity, radio luminosity, and FIR-radio flux ratio parameter $q$. Mean values with their standard deviations in parentheses and median values of the parameters are presented for the subsamples containing AGNs and without them (non-AGN) as well as for the subsample of OHMs detected in the Arecibo survey.}
\centering
\resizebox{\textwidth}{!}{%
\begin{tabular}{ccccccccccccccc}
\hline
\multirow{2}{*}{Sample} & \multirow{2}{*}{Subsample} & \multirow{2}{*}{N}  & \multicolumn{2}{c}{\multirow{2}{*}{$\alpha_{1.4}$}} & \multicolumn{2}{c}{\multirow{2}{*}{$\alpha_{4.7}$}} & \multicolumn{2}{c}{\multirow{2}{*}{$\log~R$}} & \multicolumn{2}{c}{$\log~L_{\rm FIR}$,} & \multicolumn{2}{c}{$\log~P_{1.4}$,} & \multicolumn{2}{c}{\multirow{2}{*}{$q$}} \\ 
       & & & \multicolumn{2}{c}{} & \multicolumn{2}{c}{} & \multicolumn{2}{c}{} & \multicolumn{2}{c}{$L_{\sun}$} & \multicolumn{2}{c}{W Hz$^{-1}$} & \multicolumn{2}{c}{} \\
\hline     
\multirow{3}{*}{OHM} & AGN   & 35 & -0.45 (0.39) & -0.49 & -0.52 (0.42) & -0.66 & 2.90 (1.01) & 2.88 & 11.55 (0.58) & 11.64 & 23.29 (0.94) & 23.24 & 2.29 (0.63) & 2.46 \\ 
                     & non-AGN   & 39 & -0.47 (0.35) & -0.59 & -0.54 (0.35) & -0.54 & 2.40 (0.66) & 2.46 & 11.62 (0.46) & 11.77 & 23.24 (0.59) & 23.39 & 2.42 (0.30) & 2.49  \\
                     & all & 74 & -0.46 (0.37) & -0.55 & -0.54 (0.38) & -0.59 & 2.64 (0.87) & 2.70 & 11.59 (0.52) & 11.69 & 23.26 (0.77) & 23.3 & 2.36 (0.48) & 2.48 \\ 
                     & Arecibo & 21 & -0.40 (0.49) & -0.59 & -0.44 (0.53) & -0.52 & 2.58 (0.58) & 2.63 & 11.94 (0.24) & 11.95 & 23.73 (0.28) & 23.67 & 2.28 (0.26) & 2.32 \\
\hline
\multirow{3}{*}{control} & AGN   & 38  & -0.65 (0.46) & -0.70 & -0.79 (0.44) & -0.81 & 2.31 (1.11) & 2.08 & 11.85 (0.34) & 11.82 & 24.37 (1.10) & 24.09 & 1.59 (0.98) & 2.02  \\ 
                         & non-AGN   & 90  & -0.63 (0.52) & -0.69 & -0.69 (0.44) & -0.70 & 2.08 (0.79) & 1.87 & 11.71 (0.29) & 11.66 & 23.62 (0.35) & 23.58 & 2.18 (0.23) & 2.22 \\ 
                         & all & 128 & -0.64 (0.50) & -0.70 & -0.72 (0.43) & -0.71 & 2.15 (0.9) & 1.92 & 11.75 (0.31) & 11.71 & 23.85 (0.75) & 23.67 & 2.01 (0.63) & 2.14 \\
\hline
\end{tabular}
}
\end{table*}

\subsection{Radio luminosity and radio loudness}
\label{loudness}
\begin{figure}
\centerline{\includegraphics[width=.4\textwidth]{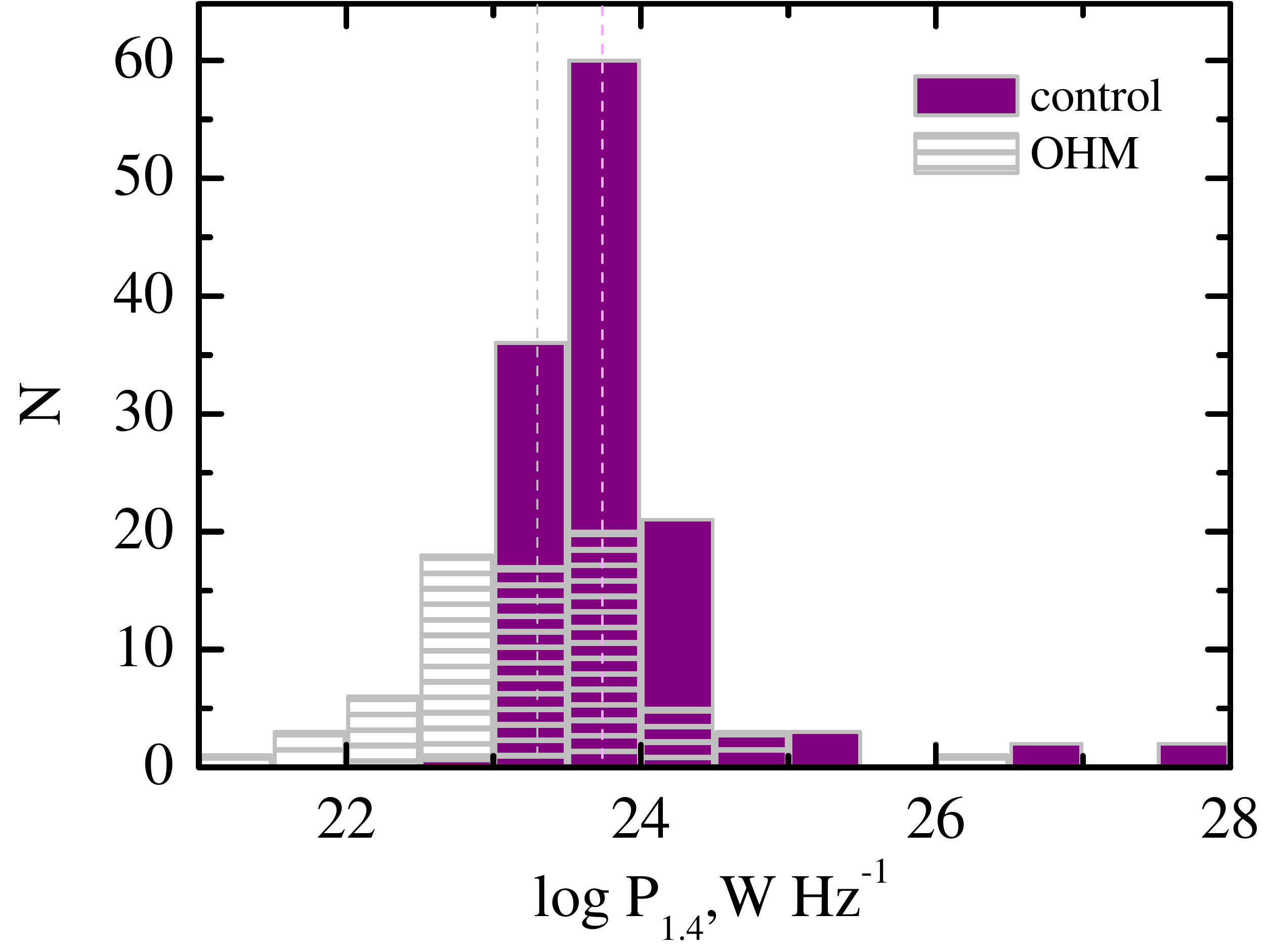}}
\caption{Distributions of the radio luminosity at 1.4 GHz
for the control (purple) and OHM (grey) samples. The grey and pink lines are, respectively, the median values of the $\log~P_{1.4}$ for the samples.}
\label{fig:logl}
\end{figure}

We used the $\Lambda$CDM cosmology with $H_0 = 67.74$~km\,s$^{-1}$\,Mpc$^{-1}$, $\Omega_m=0.3089$, 
and $\Omega_\Lambda=0.6911$ \citep{2016A&A...594A..13P} to estimate the monochromatic radio luminosity at 1.4 GHz using the formula:
\begin{equation}
P_{\nu} = 4 \pi D_{L}^2 S_{\nu} (1+z)^{-\alpha -1},
\end{equation}
where $S_{\nu}$ is the measured flux density at
a frequency $\nu$, is the redshift, $\alpha$ is the spectral index, and $D_{L}$ is the luminosity distance. The median values of $P_{1.4}$ are $\sim 2\times 10^{23}$~W\,Hz$^{-1}$ and \mbox{$\sim 5\times 10^{23}$~W\,Hz$^{-1}$} for the OHM and control samples, respectively
(Fig.~\ref{fig:logl}).

The radio loudness is defined as
\begin{equation}
R = \frac{S_{\nu, {\rm radio}}}{S_{\nu, {\rm opt}}},
\end{equation}
where $S_{\nu,{\rm radio}}$ is the radio flux density at 1.4 GHz and $S_{\nu, {\rm opt}}$ is the optical flux density corresponding to 
the $B$ or $g$ filter. We adopted $g$-band (4866~\AA) magnitudes for 71 of 74 OHMs from Pan-STARRS \citep{2016arXiv161205560C}, and $B$-band (4410~\AA) magnitudes for two sources from NOMAD \citep{2004AAS...205.4815Z}, and for one from APASS \citep{2015AAS...22533616H}. For the control sample $g$-band magnitudes are from SDSS DR12 (4770~\AA) \citep{2015ApJS..219...12A} and Pan-STARRS. The magnitudes were transformed into flux densities using the Pogson law. We adopted $S_{\nu,{\rm opt}} = 3631$ Jy for $g$ = 0,\!\footnote{https://www.sdss.org/dr12/algorithms/fluxcal/\#SDSStoAB} $S_{\nu,{\rm opt}} = 4260$ Jy for $B = 0$ \citep{1979PASP...91..589B}, and the optical spectral index $\alpha=-0.3$ \citep{2019A&A...630A.110G} for the K correction.

\begin{figure}
\centerline{\includegraphics[width=.4\textwidth]{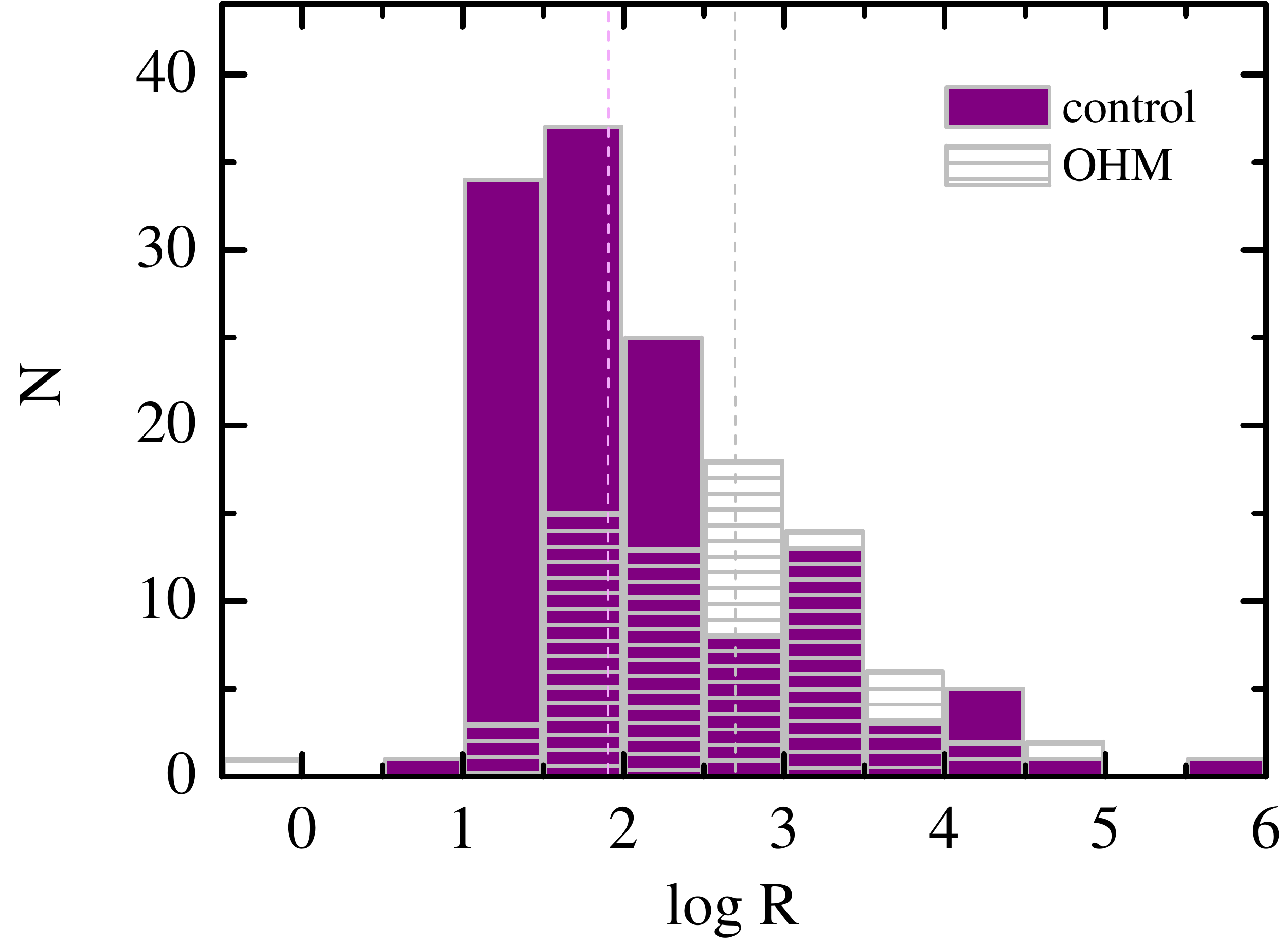}}
\caption{Radio loudness distributions for the control (purple) and OHM (grey) samples. The dashed lines show the median values of $\log$~R.}
\label{fig:logr}
\end{figure}

The median values of $\log R$ are 2.7 and 1.9 for the OHM and control samples respectively (Fig.~\ref{fig:logr}). If we adopt the radio loudness criteria similar to the one of the most common
used for quasars by \cite{1989AJ.....98.1195K}, all except one OHM (J0138-10) and all except one source in the control sample (J1337+24) are radio-loud with $\log~R > 1$. Furthermore, 57 per cent of the OHMs and 24 per cent of the sources in the control sample are highly radio-loud with $\log~R > 2.5$ \citep{2019MNRAS.482.2016Z}. 
According to the Kolmogorov--Smirnov test, both $P_{1.4}$ and $R$ distributions are different for the two samples at a significance level of 0.05.

\subsection{Emitting region properties in radio and FIR}
\label{FIR}

\begin{figure}
\centerline{\includegraphics[width=.4\textwidth]{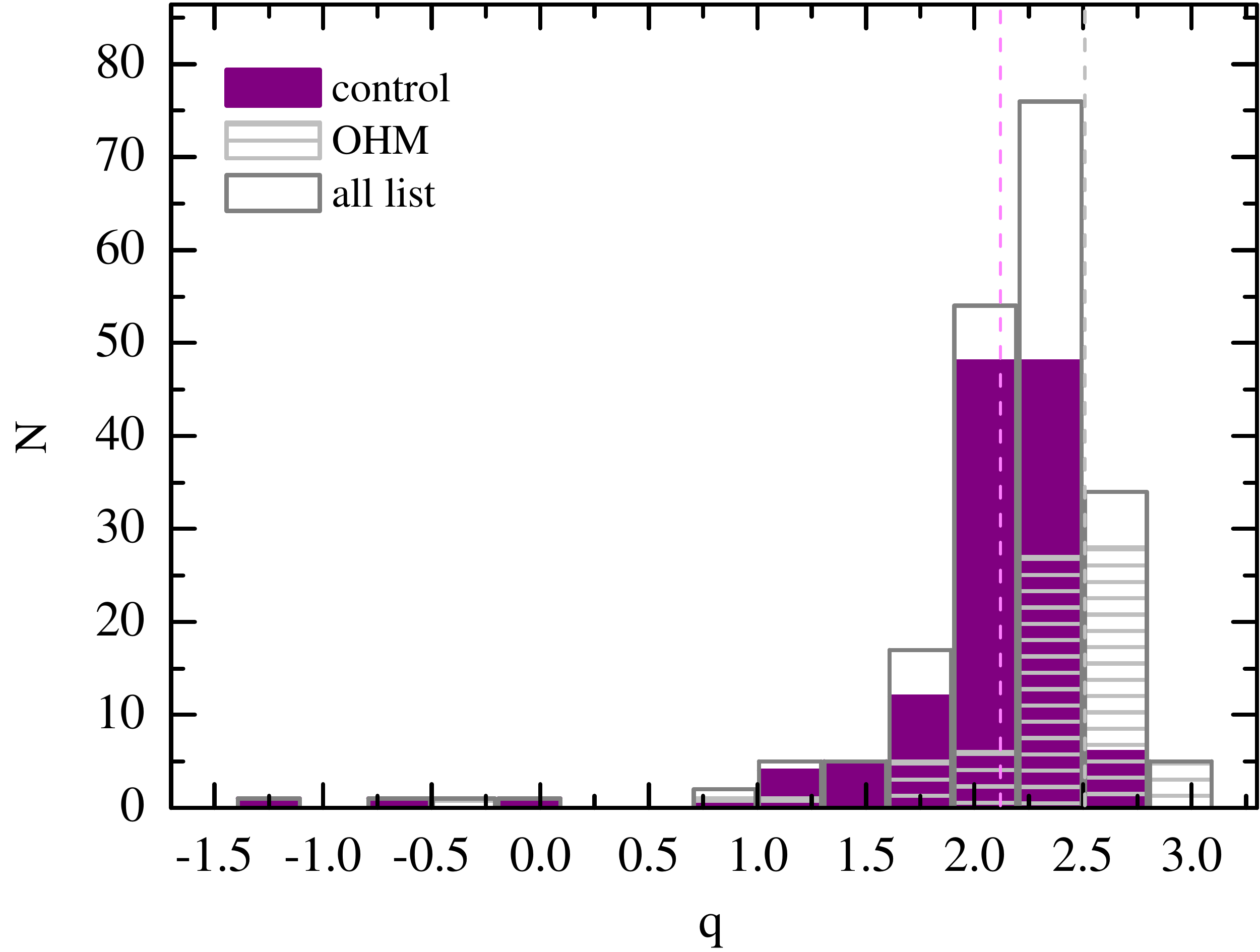}}
\caption{The $q$ ratio distributions for the 74 OHMs (grey) and 128 control sample galaxies (purple); for the entire sample the distribution of the $q$ parameter is shown by the dark grey histogram (the normal distribution at the 0.05 level). The median values of $q$ for the two samples are marked with the dashed grey and pink lines.} 
\label{fig:q}
\end{figure}

The efficiency of a galaxy in producing FIR emission compared to the nuclear radio continuum is defined by the logarithm of the FIR and radio flux density ratio or the $q$ ratio. This parameter indicates the underlying energy source nature and is determined from the equation \citep{1985ApJ...298L...7H}: 
\begin{equation}
q=\log \left [\frac{({\rm FIR}/3.75\times 10^{12})}{S_{1.4}}\right],
\end{equation}
where $S_{1.4}$ is the flux density at 1.4 GHz and $FIR$ is the far-infrared flux estimate between 40 $\mu$m and 120 $\mu$m (in units of W m$^{-2}$). $FIR$ is defined by \citep{1985ApJ...298L...7H}:
\begin{equation}
FIR = 1.26\times 10^{-14}(2.58~S_{\rm 60\mu m}+S_{\rm 100\mu m}), 
\end{equation}
where $S_{\rm 60\mu m} $ and $S_{\rm 100\mu m}$ are flux densities (Jy) at 60 and 100 $\mu m$, respectively. The median values of $q$ equal to 2.48 and 2.15 for the OHM and control samples, respectively.
The distribution of the $q$ ratio is normal (at the 0.05 level according to the Shapiro--Wilk test) for the entire sample and for each list separately (Fig.~\ref{fig:q}).

\begin{figure}
\centerline{\includegraphics[width=.45\textwidth]{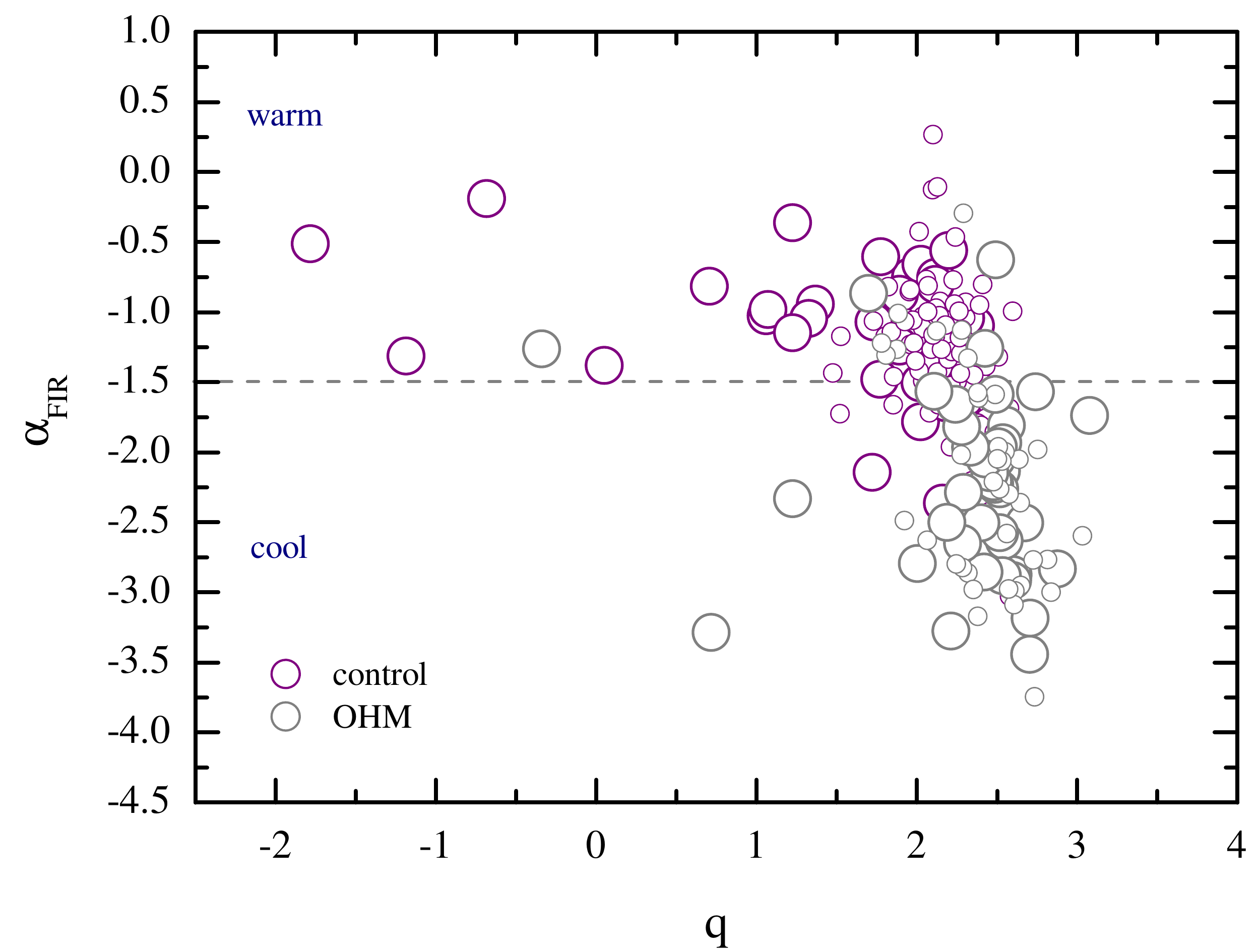}}
\begin{small}
\caption{$\alpha_{\rm FIR}$ versus $q$ ratio 
for the 74 OHMs (grey) and 128 control sample galaxies (purple). The dashed horizontal line separates the regions of ``warm'' ($\alpha_{\rm FIR} > -1.5$) and ``cool'' ($\alpha_{\rm FIR} < -1.5$) IRAS galaxies \citep{1992A&AS...96..389D,2007MNRAS.375..931M}. AGNs are marked with the larger circles. 
}
\label{fig:qfir}
\end{small}
\end{figure}

The FIR spectral index, measured between 25 and 60$\mu$m as:
\begin{equation}
\alpha_{\rm FIR} = \frac{\log (S_{\rm 25\mu m} / S_{\rm 60\mu m})} {\log (60/25)}
\end{equation}
is parameter, which describes the temperature of the FIR emitting region. SF galaxies have cooler dust emission and the values of $\alpha_{\rm FIR}<-1.5$. The central engines of AGNs heat the dust to warmer temperatures and have $\alpha_{\rm FIR}>-1.5$ \citep{1992A&AS...96..389D,2007MNRAS.375..931M}. We plot the $\alpha_{\rm FIR}$ versus $q$ ratio in Fig.~\ref{fig:qfir}, where the dashed horizontal line separates the regions of ``warm'' ($\alpha_{\rm FIR} > -1.5$) and ``cool'' ($\alpha_{\rm FIR} < -1.5$) galaxies. Interesting to note that only 4 per cent of AGNs in OHM sample have ``warm'' dust emission. In the control sample 82 per cent of AGN belong to ``warm''. In general, only 16 per cent of OHMs and 72 per cent of the control sample in a ``warm'' area.

There are several proposed values of the $q$ parameter for separation of SF galaxies and AGNs. The radio-excess and IR-excess galaxy limits $q\leq1.64$ and $q\geq3.05$, respectively, are taken by \cite{2001ApJ...554..803Y}. In the papers of \cite{1991AJ....102.1663C} and \cite{2001ApJ...554..803Y}, a mean value $q=2.34$ was found as typical in SF galaxies. A cut-off of $q=1.8$ is used as a diagnostic tool to distinguish between SF galaxies and radio-loud AGNs in \cite{2002AJ....124..675C}. The mean $q$ ratio for our OHM sample is 2.36 with an rms scatter 
$\sigma_{q}=0.48$, which is close to $q=2.34$ ($\sigma_{q}=0.01$) that was found for SF galaxies in \cite{2001ApJ...554..803Y}.
In Table~\ref{tab:qhigh}, we list the sources with $q\leq1.64$: there are three OHM AGNs (all of them are radio-loud, with $\log R \geq  2.5$) and 14 galaxies in the control sample, 12 out of which are radio-loud and 11 are AGNs. The control sample contains more radio-loud and radio-excess AGNs (11 per cent) than the OHM sample (5 per cent). Most radio-excess sources in this list (12 non-OHMs, and one OHM) have steep spectra.

The correlation between radio and FIR luminosities is common for the galaxies with star formation, while AGNs deviate from this correlation making it an important diagnostic tool to separate SF galaxies and radio-loud AGNs. \cite{2007MNRAS.375..931M} found the following relationship for a large sample of SF galaxies:
\begin{equation}
\label{eq:mauch2007}
\log (P_{1.4}) = (1.06\pm0.01) \log (L_{\rm FIR}) + (11.1\pm0.1).
\end{equation}

We used the data from the IRAS Point Source Catalog v2.1 and Faint Source Catalog v2.0\footnote{\url{https://irsa.ipac.caltech.edu/applications/Gator/}} \citep{1990BAAS...22Q1325M,1990IRASF.C......0M} to calculate the far-infrared luminosity $L_{FIR}$ as:
\begin{equation}
L_{FIR} = 4 \pi D_{L}^2 FIR (1+z)^{-\alpha_{IR} -1},
\end{equation}
where $\alpha_{IR}$ is the infrared spectral index. Obtained values of $L_{\rm FIR}$ show that the hosts of OHMs are mostly LIRGs ($L_{\rm FIR} >$ 10$^{11}L_{\sun}$, 62/74), and about one-fifth of them are ULIRGs ($L_{\rm FIR} >$ 10$^{12}L_{\sun}$, 15/74). In a control sample 127/128 are LIRGs and 26/128 (also one-fifth) are ULIRGs.

According to the Kolmogorov--Smirnov test, these parameters ($L_{\rm FIR}$, $q$, and $\alpha_{\rm FIR}$) distributions for the two samples are significantly different at 0.05 level. 

We found a correlation between $L_{\rm FIR}$ and $P_{1.4}$ with the Spearman $\rho$=0.82$_{-0.86}^{+0.76}$ (p-val.$\sim 10^{-19}$) for the OHM sample and $\rho$=0.62$_{-0.68}^{+0.55}$ (p-val.$\sim 10^{-15}$) for the control sample. These values are the median and range of the 16 and 84 percentiles computed with 1000 bootstraps using the \textsc{pymccorrelation} package\footnote{\url{https://github.com/privong/pymccorrelation}}. We then used the partial correlation coefficient method \citep{1982MNRAS.199.1119M,1996Ap.....39..237K} to take into account the Malmquist effect and found that a correlation retains significant with $\rho$=0.55 (p-val.$\sim 5\times 10^{-7}$) for the OHM sample, but becomes insignificant for the control sample with $\rho$=0.14 (p-val.=0.11).

\begin{figure}
\centerline{\includegraphics[width=.40\textwidth]{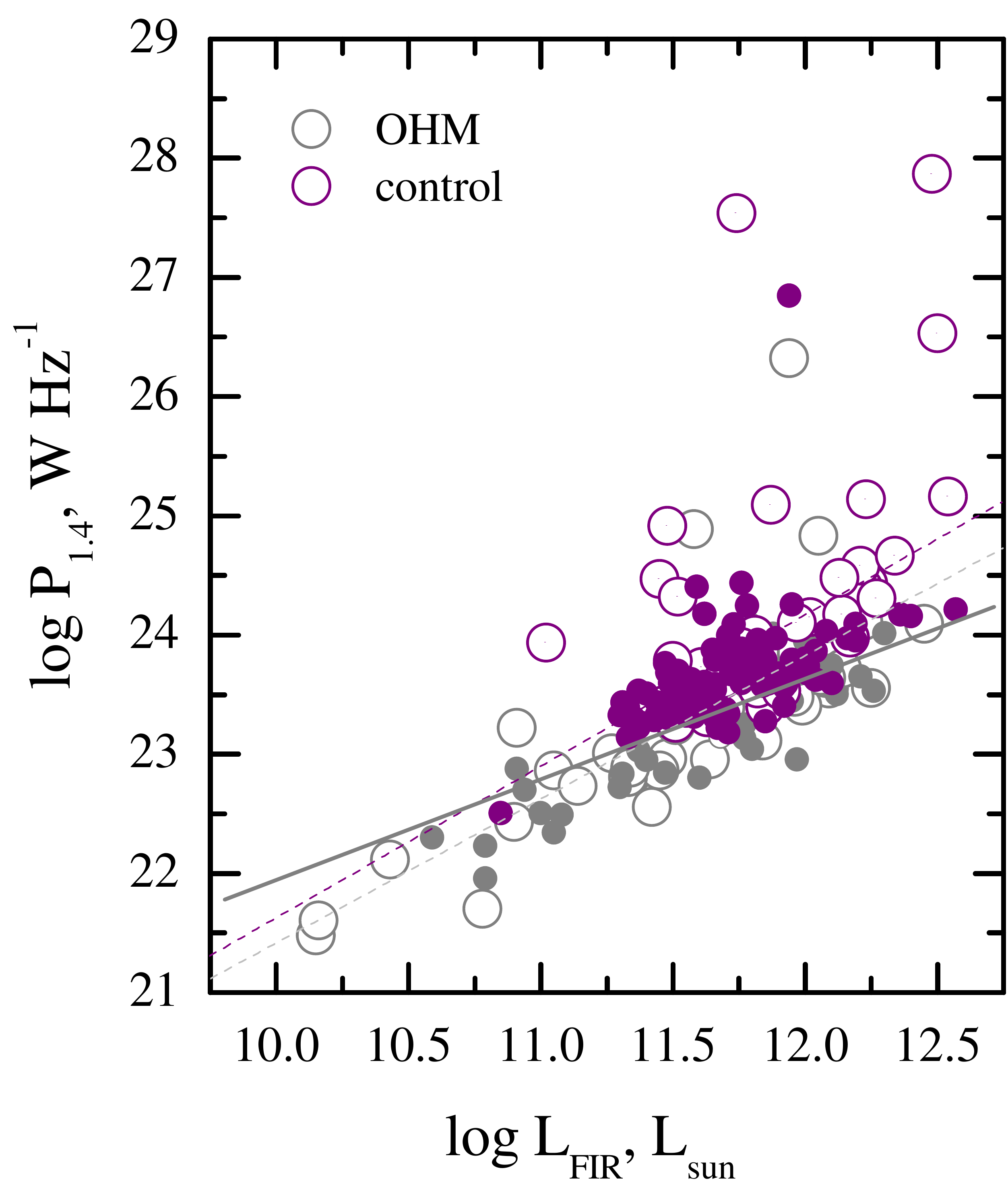}}
\caption{The relation between the radio luminosity at 1.4 GHz and the FIR luminosity for 74 OHM galaxies (grey) and 128 control sample galaxies (purple). AGNs are marked with the open circles. A linear fit is shown with the grey dashed line for the OHMs and with the purple dashed line for the control sample. The solid grey line is a linear fit for the OHMs taking into account the Malmquist effect.}
\label{fig:firradio}
\end{figure}

The relation between the radio luminosity for 1.4 GHz and the FIR luminosity is presented in Fig.~\ref{fig:firradio}. A least-squares line of 
the best fit to the $L_{\rm FIR}$--$P_{1.4}$ relation for the OHM sample, 
allowing for the Malmquist bias, is described by
\begin{equation}
\log (P_{1.4}) = (0.82\pm 0.08) \log (L_{\rm FIR}) + (13.81\pm 0.96) 
\end{equation} 
and by
\begin{equation}
\log (P_{1.4}) = (0.35\pm0.03) \log (L_{\rm FIR}) + (19.79\pm0.41)
\end{equation} 
for the control sample. 
We calculated smaller slopes compared to the results of \cite{1991ApJ...376...95C} (1.11), \cite{2001ApJ...554..803Y} (0.99$\pm$0.01), and \cite{2007MNRAS.375..931M} (1.06$\pm$0.01). Compared to mentioned studies, our values obtained taking into account the Malmquist bias and also our samples are brighter in the radio domain (up to 10$^{26}$ W Hz$^{-1}$), containing a large fraction of AGNs (up to 47 per cent), which might cause the bigger scatter in radio-FIR plot \citep{2001ApJ...554..803Y} and higher dispersion in measured coefficients of the equations.

\begin{table}
 \caption{\label{tab:qhigh} List of the radio-excess galaxies with $q\leq1.64$.}
 \centering
 \begin{tabular}{lcccc}
 \hline
 \multirow{2}{*}{Name} & \multirow{2}{*}{Type} & \multirow{2}{*}{$q$} & \multirow{2}{*}{$\log~R$} & \multirow{2}{*}{sp. type} \\ 
      &      &   &        &  \\ 
\hline
\multicolumn{5}{c}{OHM} \\
\hline
J0251+43 & AGN &  0.72 & 4.9 & peaked\\ 
J0159-29 & AGN &  1.21 & 4.1 & steep \\ 
J1347+12 & AGN & -0.33 & 4.8 & flat \\ 
\hline
\multicolumn{5}{c}{control} \\  
\hline
J0137+33 &    AGN & -1.18 & 4.2 & steep \\ 
J0153+26 &    AGN &  1.37 & 2.2 & steep \\ 
J0817+31 &    AGN &  1.06 & 5.7 & steep \\ 
J0854+20 &    AGN &  0.05 & 2.9 & rising \\ 
J0945+17 &    AGN &  1.33 & 2.2 & steep \\ 
J1114+32 &    AGN &  1.07 & 3.1 & ultra-steep \\ 
J1229+02 &    AGN & -1.78 & 3.4 & peaked \\ 
J1356+10 &    AGN &  1.23 & 1.9 & steep \\ 
J1614+26 &    AGN &  1.22 & 1.2 & steep \\ 
J1751+26 &    AGN &  0.71 & 3.2 & steep \\ 
J1817+15 & non-AGN &  1.53 & 2.4 & steep \\ 
J1958+16 & non-AGN &  1.48 & 4.6 & steep \\ 
J2313+03 & non-AGN &  1.52 & 3.2 & steep \\ 
J2316+04 &    AGN & -0.68 & 4.3 & steep \\ 
\hline
\end{tabular}
\end{table}

\begin{figure}
\centerline{\includegraphics[width=.4\textwidth]{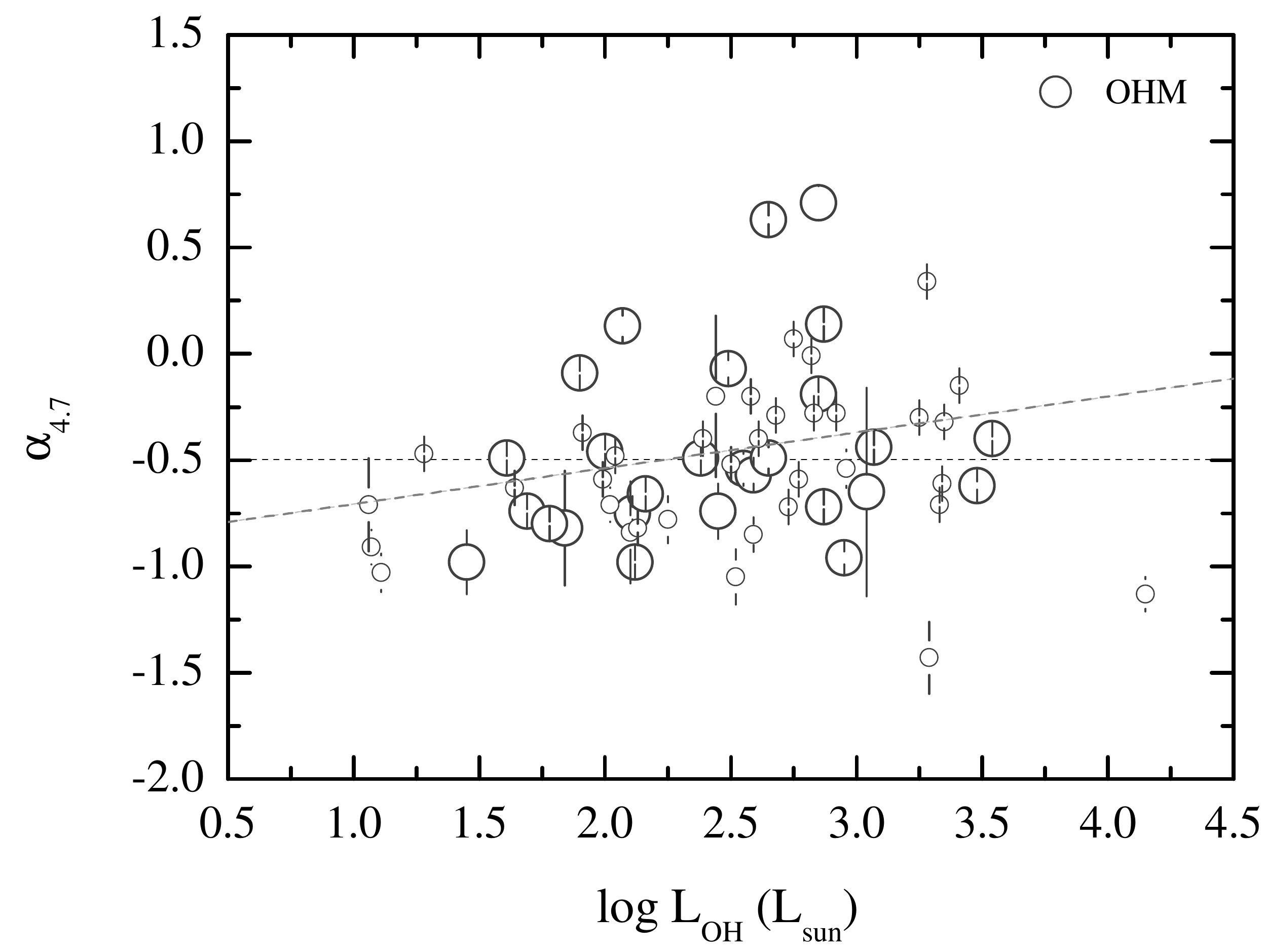}}
\caption{ $\log L_{\rm OH}$ versus $\alpha_{4.7}$ for 62 OHM galaxies.
AGNs are marked with larger circles. The linear fit is shown with the grey line. The area of flat and rising spectra $\alpha \geq -0.5$ is shown with the black dotted line.}
\label{fig:oha}
\end{figure}

For the OHM sample, we checked the relation between the radio spectrum
type and isotropic OH line luminosity $L_{OH}$ taken from \cite{2014A&A...570A.110Z}, where it is available for 62 sources.
In Fig.~\ref{fig:oha} the AGNs are shown by the larger circles, the black dotted line corresponds to the area of $\alpha\geq-0.5$ (flat and rising radio spectra). We found a correlation between $\alpha_{4.7}$ and $L_{OH}$ with the Spearman $\rho$=0.26$_{-0.38}^{+0.14}$ (p-val.=0.04$_{-0.29}^{+0.002}$). The correlation is getting stronger if two outliers are excluded from the diagram ($\rho$=0.38$_{-0.48}^{+0.26}$, p-val.=0.003$_{-0.04}^{+0.0001}$).  
We also discovered a correlation between $L_{OH}$ and $P_{1.4}$ ($\rho$=0.66$_{-0.74}^{+0.56}$, p-val.=7$\times$10$^{-9}$) for OHMs, and it decreased but remained significant with $\rho$=0.35 (p-val.=0.005) after considering the Malmquist effect.

\section{Discussion}

Steep spectra at the frequencies below $\sim$10 GHz ($\alpha\sim-0.8$) is a common feature of galaxies with starburst (e.g. \citealt{1981A&A....94...29K,2019ApJ...875...80G,2021MNRAS.507.2643A}), but it is also common in AGNs having radio lobes with their emitting material cooled down by synchrotron aging. However, even flat spectra cannot rule out starburst activity in FIR galaxies \citep{1991ApJ...378...65C,1997A&A...322...19N,2006A&A...449..559B}.
\citet{2005Ap.....48..237K} found on average flat ($\alpha=-0.36\pm0.04$) spectral indices, measured between 1.49 and 5/8.44 GHz, for the OHMs powered by both starburst activity and AGNs. The author also reported that among the AGN and starburst galaxies of his sample, the fractions of galaxies with a flat or steep spectrum are almost the same; however, noticing that the number of galaxies in each subsample is not enough for reliable statistics. Our results on radio spectrum types among AGN and non-AGN OHMs do not contradict 
those of \cite{2005Ap.....48..237K}. 29 of the 30 OHMs from the \cite{2005Ap.....48..237K} list were observed with RATAN-600 among the 74 OHMs in our sample. For 22 OHMs \cite{2005Ap.....48..237K} had presented spectral indices, and we compared them with our measured $\alpha_{4.7}$ indices. They do linearly correlate (the Pearson $r=0.44_{-0.64}^{+0.21}$, p-val.=0.04$_{-0.34}^{+0.001}$), but our values show systematically steeper spectra for 17 sources of 22, which is confirmed by the average spectral indices for these 22 sources: $\alpha_{4.7}=-0.53$ (0.32) by our measurements and $\alpha=-0.36$ (0.19) in \cite{2005Ap.....48..237K}. 
This could be explained by higher resolution of the VLA observations where
they resolve the central part of a source with the flat spectrum, while
with RATAN-600 we measure the total flux density from a more extensive region. The fluxes in the OH line and radio continuum at 1.49 GHz are found to be uncorrelated ($\rho$=0.32, p-val.=0.07) for the 87 galaxies studied in \citet{2005Ap.....48..237K}. In contrast to that, we detect a correlation between $L_{OH}$ and $P_{1.4}$ luminosities ($\rho$=0.35, p-val.=0.005) in our OHM sample.

In our samples, both at 1.4 and 4.7 GHz, the majority of the sources are classified as having a steep spectrum, but among OHMs significant part (32 per cent) of the sample have flat spectra ($-0.5 \leq \alpha \leq 0$). A third of analysed radio spectra have measurements at only two frequencies and many of them were measured non-simultaneously. All this influences the obtained spectral classification and, in fact, part of the sources with flat spectra could have steep spectra. 

Radio and FIR properties of U/LIRGs galaxies are an effective instrument for investigating the dominant energy source in their central regions, whether it is extensive star formation or radiation from the accreting supermassive black hole in an AGN. We confirm the tight correlation of the FIR and radio luminosities 
for the OHM galaxies \citep{1991ApJ...376...95C,2007MNRAS.375..931M}. The FIR/radio flux ratios, measured by the parameter $q$, indicate a difference between the galaxies with OH emission and the galaxies in the control sample: it is caused by the difference in radio power, the median value of which is two times less for the OHMs than that for the control sample (Table~\ref{tab:stats}). 
The sources from the control sample are located further (at $z$ up to 0.37) and spread at a wider range of radio power. The $q$ ratio
corresponds to the total energetic contribution of the radio and infrared emission in U/LIRGs \citep{1985ApJ...298L...7H,1991AJ....102.1663C,1991AJ....101..362C,2001ApJ...554..803Y}. Its continuous distribution for the OHM and non-OHM galaxies can reflect coexistence of AGN-dominated and starburst-dominated sources in U/LIRGs. 

The obtained correlations between independent parameters of the spectral index $\alpha_{4.7}$ and the OH line luminosity $L_{OH}$, and also between the 1.4 GHz radio luminosity $P_{1.4}$ and the OH line luminosity $L_{OH}$, could be an indicator of the indirect connection between a bright radio core and the production of the OH emission line. 

A significant part of both samples are galaxies classified as an AGN,
which could explain the similarities in the radio properties we determined, such as the spectral index and radio loudness.
Table~\ref{tab:stats} shows the statistics for the subsamples of 
the OHM and control samples with and without AGNs. For the OHM subsample with excluded AGNs, standard deviations of parameters are always less than for the AGN-containing OHM subsample. 
For the control sample standard deviations also appear to be less for the subsample with excluded AGNs for all parameters, except the spectral indices. 

The two samples under investigation are compiled using different approaches. While the control sample is exclusively collected from the non-detections of the Arecibo survey, the OHM sample is a compilation of all the known megamaser galaxies, taking into account the same flux density limit. We therefore examined the OH detected megamasers subsample selected from the Arecibo survey \citep{2002AJ....124..100D}. The OHMs with $S_{1.4}> 5$ mJy from the Arecibo survey are represented in our study by 21 objects. We determined their spectral types mostly as steep (10 galaxies) and flat (6), also there are two objects of the rising type and one representative of the peaked, upturn, and ultra-steep types. Their mean spectral indices at 1.4 and 4.7 GHz are both flat, $-$0.40 (0.49) and $-$0.44 (0.53), which is consistent with the values for the whole OHM sample, although being insignificantly flatter (Table~\ref{tab:stats}).

Our samples are selected from \cite{2014A&A...570A.110Z}, and 
we can compare our conclusions based on the radio and MID properties of the OHMs. First, the AGN fraction, calculated using the WISE magnitude-based AGN 
criterion $W1-W2 \geq 0.8$, is similar for both the OHM and non-OHM samples (about 40 per cent) in \cite{2014A&A...570A.110Z}. In our OHM and non-OHM samples, the AGN fraction is 47 and 30 per cent, respectively. 
This is a good match between the MIR AGN defining criterion and the AGN catalogue classification, the difference is likely caused by our flux density selection and therefore limited samples. Secondly, \cite{2014A&A...570A.110Z} show
that the distribution of the MIR spectral index $\alpha_{22-12\mu m}$ is significantly different for the OHM and non-OHM samples. We revealed that 
the distributions of the radio spectral indices are also significantly different for the two samples. Thirdly, the correlation of the MIR luminosity and the maser luminosity was revealed to be marginal in \cite{2014A&A...570A.110Z}. We found that continuum radio emission at 1.4 GHz correlates with maser luminosity, although being not so strong as the correlation between maser and FIR luminosities. 
These results are consistent, supporting the assumptions about OH emission being pumped by lower energy photons of FIR/radio emission.

\section{Summary}

We present the radio properties of OHM galaxies by analysing radio continuum spectra of two U/LIRGs samples: the OHM sample (74 objects) and the non-OHM control sample (128 objects).
The new multifrequency RATAN-600 measurements and available literature radio data were used for compilation of radio continuum spectra. We have revealed a low detection rate at the RATAN frequencies (4--75 per cent). In general, the detection rate is higher for the OHMs than 
for the control sample. We measured the radio continuum spectra for 57 per cent of objects from both samples at frequencies higher than 4.7 GHz for the first time. The main conclusions are the following:

\begin{enumerate}
\item The samples stand out statistically by the flat-spectrum fraction. The OHM sample has 32 per cent sources with flat spectra,
while the control sample has 18 per cent. Steep radio spectra prevail in the samples: 53 per cent and 61 per cent for the OHM and control samples respectively. We determined the median spectral indices at 1.4 GHz as $-0.55$ and $-0.70$, and at 4.7 GHz as $-0.59$ and $-0.71$ for the OHM and control lists, respectively. The distributions of radio parameters such as the spectral index, radio loudness, and radio luminosity are different at a significance level of 0.05 according to the Kolmogorov--Smirnov test for the control and OHM samples.
\item We have found a weak correlation between $\alpha_{4.7}$ and isotropic OH line luminosity $L_{OH}$ ($\rho$=0.26 and p-val.=0.04) 
as well as a significant correlation between $L_{OH}$ and radio luminosity $P_{1.4}$ ($\rho$=0.35 and p-val.=0.005) taking into account the Malmquist effect. These results suggest a substantial role of the radio continuum emission in producing the OH line emission in masers.
\item The $q$ ratio parameter has a normal distribution (at the 0.05 level) for the two samples and reflects a continuous distribution ranging between AGN-dominated and SF-dominated sources in U/LIRGs. This confirms the general assumption about coexistence of AGNs and starbursts in U/LIRGs \citep{2006A&A...449..559B}. The distributions of $q$ ratio for the OHM and control samples are significantly different.
\end{enumerate}

The obtained complex relations between radio, OH line luminosity, and FIR parameters reflect the influence of various properties of AGN and SF objects in the samples.

\section*{Acknowledgements}
We thank the referee for providing suggestions on results of previous studies on topic and useful comments that improved the paper.
The reported study was funded by RFBR and NSFC, project number 21-52-53035 ``The Radio Properties and Structure of OH Megamaser Galaxies''. The Chinese team acknowledges the support from the NSFC (No. 11763002, No. U1931203, No. 12111530009). The observations were carried out with the RATAN-600 scientific facility. Observations with RATAN-600 are supported by the Ministry of Science and Higher Education of the Russian Federation. 
This paper has been supported by the Kazan Federal University Strategic Academic Leadership Program (``PRIORITY-2030").

This research has made use of the NASA/IPAC Extragalactic Database (NED), which is operated by the Jet Propulsion Laboratory, California Institute of Technology, under contract with the National Aeronautics and Space Administration; the CATS data base, available at the Special Astrophysical Observatory website; the SIMBAD data base, operated at CDS, Strasbourg, France; \textsc{pymccorrelation} package, which implements the Monte Carlo error analysis procedure described in \cite{2014arXiv1411.3816C}, first used in \cite{2020ApJ...893..149P}; NumPy \citep{2020NumPy-Array}; SciPy \citep{2020SciPy-NMeth}.

\section*{Data Availability}
The data underlying this article are available in the article itself and in its online supplementary material. The measured flux densities
are distributed in the VizieR Information System.

\bibliographystyle{mnras} %
\bibliography{manuscript} %

\newpage
\onecolumn 
\appendix
\section{Sample parameters}
\label{sec:tables}

\begin{small}
\begin{longtable}{|l|l|c|c|c|c|c|c|c|}
\caption{\label{tab:OHM}The OHM sample parameters: the right ascension R.A. and declination Dec, redshift z, flux density at 1.4 GHz $S_{1.4}$, radio luminosity $P_{1.4}$ and radio loudness $\log~R$, spectral indices at 1.4 GHz $\alpha_{1.4}$ and at 4.7 GHz $\alpha_{4.7}$, type of radio spectrum.}\\
\hline
R.A.	&	Dec.	&	$z$	&	$S_{1.4}$, Jy	&	$P_{1.4}, 10^{23}$ W Hz$^{-1}$ &	$\log~R$	&			$\alpha_{1.4}$	&	$\alpha_{4.7}$	&	type	\\
\hline
1 & 2 & 3 & 4 & 5 & 6 & 7 & 8 & 9  \\
\hline
\endfirsthead
\caption[]{continued.}\\
\hline
R.A.	&	Dec	&	$z$	&	$S_{1.4}$, Jy	&	$P_{1.4}, 10^{23}$ W Hz$^{-1}$	&	$\log~R$	&			$\alpha_{1.4}$	&	$\alpha_{4.7}$	&	type	\\
\hline
1 & 2 & 3 & 4 & 5 & 6 & 7 & 8 & 9 \\
\hline
\endhead
\hline
\endfoot
00:08:20.86	&	$+$40:37:55.3	&	0.0447	&	0.007$\pm$0.001	&	0.36$\pm$0.02	&	1.84$_{-0.04}^{+0.03}$	&	-0.09$\pm$0.01	&	-0.09$\pm$0.01	& flat \\
00:36:00.29	&	$-$27:15:34.9	&	0.0693	&	0.011$\pm$0.001	&	1.35$\pm$0.06	&	1.67$\pm$0.03	&	-0.40$\pm$0.01	&	-0.40$\pm$0.01	& flat \\
00:47:33.13	&	$-$25:17:17.1	&	0.0008	&	2.995$\pm$0.113 &	0.04$\pm$0.002	&	3.38$\pm$0.03	&	-0.70$\pm$0.01	&	-0.70$\pm$0.01	& steep\\ 
00:53:34.77	&	$+$12:41:33.8	&	0.0589	&	0.009$\pm$0.001	&	0.77$\pm$0.04	&	1.76$\pm$0.03	&	-0.80$\pm$0.01	&	-0.80$\pm$0.01	& steep	\\
01:38:52.87	&	$-$10:27:11.7	&	0.0482	&	0.016$\pm$0.001	&	0.90$\pm$0.05	&	-0.43$\pm$0.03	&	-0.46$\pm$0.01	&	-0.46$\pm$0.01	& flat\\
01:44:30.55	&	$+$17:06:08.0	&	0.0274	&	0.041$\pm$0.001	&	0.73$\pm$0.01	&	2.37$\pm$0.03	&	-0.28$\pm$0.05	&	-0.50$\pm$0.05	& flat	\\
01:59:03.26	&	$+$25:42:36.8	&	0.1657	&	0.006$\pm$0.001	&	4.56$\pm$0.37	&	3.17$_{-0.08}^{+0.07}$	&	-0.30$\pm$0.01	&	-0.30$\pm$0.01	& flat \\
01:59:13.82	&	$-$29:24:35.8	&	0.1400	&	0.124$\pm$0.001 &	67.50$\pm$0.35	&	4.10$\pm$0.03	&	-0.82$\pm$0.02	&	-0.96$\pm$0.03	& steep	\\
02:42:40.72	&	$-$00:00:47.7	&	0.0038	&	4.849$\pm$0.170	&	1.65$\pm$0.06	&	4.03$\pm$0.03	&	-0.71$\pm$0.01	&	-0.85$\pm$0.02	& steep \\
02:51:34.52	&	$+$43:15:15.7	&	0.0514	&	1.225$\pm$0.001	&	77.30$\pm$0.10	&	4.94$\pm$0.02	&	 0.14$\pm$0.04	&	-0.08$\pm$0.04	& peaked \\
03:08:30.72	&	$+$20:46:20.3	&	0.0274	&	0.018$\pm$0.001	&	0.32$\pm$0.02	&	1.51$\pm$0.03	&	-0.47$\pm$0.01	&	-0.47$\pm$0.01	& flat \\
03:28:24.23	&	$-$14:12:06.8	&	0.0434	&	0.011$\pm$0.001	&	0.53$\pm$0.02	&	2.96$\pm$0.03	&	-0.59$\pm$0.01	&	-0.59$\pm$0.01	& steep \\
03:33:36.46	&	$-$36:08:25.9	&	0.0055	&	0.377$\pm$0.013	&	0.27$\pm$0.01	&	1.76$\pm$0.02	&	-0.68$\pm$0.01	&	-0.68$\pm$0.01	& steep \\
03:54:41.94	&	$+$00:37:03.2	&	0.1519	&	0.007$\pm$0.001	&	3.40$\pm$0.47	&	2.63$_{-1.00}^{+0.30}$	&	 0.92$\pm$0.27	& 	-0.26$\pm$0.38	& peaked \\
05:12:09.18	&	$-$24:21:56.2	&	0.0335	&	0.020$\pm$0.001	&	0.54$\pm$0.01	&	2.67$_{-0.05}^{+0.04}$	&	-0.15$\pm$0.01	&	 0.13$\pm$0.05	& upturn \\
05:21:01.38	&	$-$25:21:45.3	&	0.0426	&	0.029$\pm$0.001	&	1.30$\pm$0.02	&	3.61$\pm$0.03	&	-0.66$\pm$0.01	&	-0.66$\pm$0.01	& steep \\
05:45:48.02	&	$+$58:42:03.4	&	0.0149	&	0.138$\pm$0.005	&	0.73$\pm$0.03	&	3.36$_{-0.03}^{+0.02}$	&	-0.71$\pm$0.01	&	-0.71$\pm$0.01	& steep \\
06:22:22.43	&	$-$36:47:42.9	&	0.1080	&	0.033$\pm$0.001	&	10.20$\pm$0.44	&	2.14$\pm$0.03	&	-0.71$\pm$0.01	&	-0.71$\pm$0.01	& steep \\
06:51:45.85	&	$+$22:04:27.9	&	0.1433	&	0.011$\pm$0.001	&	5.55$\pm$0.26	&	2.08$\pm$0.03	&	-0.01$\pm$0.01	&	-0.01$\pm$0.01	& flat \\
08:09:47.24	&	$+$05:01:09.0	&	0.0522	&	0.036$\pm$0.001	&	2.46$\pm$0.04	&	2.87$\pm$0.05	&	-0.70$\pm$0.10	&	-0.79$\pm$0.11	& steep \\
08:47:50.07	&	$+$23:21:11.4	&	0.1515	&	0.006$\pm$0.001	&	3.96$\pm$0.34	&	1.59$\pm$0.06	&	-0.85$\pm$0.05	&	-0.85$\pm$0.05	& steep  \\
09:06:34.11	&	$+$04:51:25.5	&	0.1251	&	0.007$\pm$0.001	&	2.65$\pm$0.20	&	1.82$\pm$0.04	&	-0.28$\pm$0.01	&	-0.28$\pm$0.01	& flat \\
09:35:51.83	&	$+$61:21:12.9	&	0.0394	&	0.171$\pm$0.006	&	6.46$\pm$0.22	&	2.41$\pm$0.02	&	-0.49$\pm$0.01	&	-0.49$\pm$0.01	& flat \\
09:56:34.19	&	$+$08:43:03.8	&	0.1289	&	0.010$\pm$0.001	&	4.23$\pm$0.54	&	3.29$_{-0.08}^{+0.07}$	&	-0.62$\pm$0.02	&	-0.62$\pm$0.02	& steep \\
10:06:05.14	&	$-$33:53:17.2	&	0.0341	&	0.025$\pm$0.002	&	0.70$\pm$0.05	&	1.99$\pm$0.03	&	-0.69$\pm$0.01	&	-0.28$\pm$0.01	& steep \\
10:06:26.35	&	$+$27:25:43.7	&	0.1655	&	0.006$\pm$0.001	&	4.71$\pm$0.38	&	2.78$_{-0.07}^{+0.06}$	&	-0.52$\pm$0.01	&	-0.52$\pm$0.01	& steep \\
10:20:00.28	&	$+$08:13:35.4	&	0.0491	&	0.011$\pm$0.001	&	0.63$\pm$0.06	&	1.86$_{-0.06}^{+0.05}$	&	-0.29$\pm$0.01	&	-0.29$\pm$0.01	& flat \\
10:36:38.13	&	$+$15:32:39.5	&	0.1972	&	0.005$\pm$0.001	&	4.51$\pm$0.45	&	1.74$_{-0.06}^{+0.05}$	&	 0.63$\pm$0.02	&	 0.63$\pm$0.02	& rising \\
10:40:29.15	&	$+$10:53:20.5	&	0.1363	&	0.009$\pm$0.001	&	4.34$\pm$0.29	&	3.20$\pm$0.04	&	-0.40$\pm$0.01	&	-0.40$\pm$0.01	& flat \\
11:03:53.89	&	$+$40:50:59.9	&	0.0345	&	0.037$\pm$0.001	&	1.07$\pm$0.02	&	2.97$_{-0.04}^{+0.03}$	&	-0.60$\pm$0.07	&	-0.71$\pm$0.08	& steep \\
11:28:32.61	&	$+$58:33:46.8	&	0.0104	&	0.678$\pm$0.025	&	1.74$\pm$0.07	&	2.78$_{-0.08}^{+0.07}$	&	-0.79$\pm$0.14	&	-0.99$\pm$0.15	& steep \\
11:53:11.77	&	$-$39:07:48.3	&	0.0108	&	0.110$\pm$0.004	&	0.31$\pm$0.01	&	1.40$_{-0.03}^{+0.02}$	&	-0.63$\pm$0.01	&	-0.63$\pm$0.01	& steep \\
12:04:24.26	&	$+$19:25:12.3	&	0.1686	&	0.007$\pm$0.001	&	5.21$\pm$0.40	&	1.89$\pm$0.04	&	-0.72$\pm$0.01	&	-0.72$\pm$0.01	& steep \\
12:05:47.76	&	$+$16:51:08.5	&	0.2178	&	0.029$\pm$0.001	&	41.20$\pm$0.93	&	2.68$_{-0.04}^{+0.03}$	&	-0.84$\pm$0.06	&	-1.15$\pm$0.07	& steep \\
12:09:44.95	&	$-$05:01:16.3	&	0.1284	&	0.008$\pm$0.002	&	3.54$\pm$1.00	&	3.24$_{-0.20}^{+0.10}$	&	-0.44$\pm$0.01	&	-0.44$\pm$0.01	& flat \\
12:13:45.94	&	$+$02:48:40.8	&	0.0733	&	0.024$\pm$0.001	&	3.17$\pm$0.13	&	2.73$_{-0.06}^{+0.05}$	&	-0.31$\pm$0.08	&	-0.56$\pm$0.09	& flat \\
12:26:54.54	&	$-$00:52:38.8	&	0.0073	&	0.041$\pm$0.001	&	0.05$\pm$0.001	&	2.32$\pm$0.02	&	 0.31$\pm$0.01	&	-0.73$\pm$0.01	& peaked \\
12:56:14.13	&	$+$56:52:23.8	&	0.0422	&	0.310$\pm$0.012	&	13.10$\pm$0.51	&	3.80$\pm$0.03	&	 0.14$\pm$0.01	&	 0.14$\pm$0.01	& rising \\
13:12:26.36	&	$-$15:47:51.3	&	0.0099	&	0.084$\pm$0.003	&	0.20$\pm$0.01	&	2.46$_{-0.08}^{+0.07}$	&	-0.79$\pm$0.09	&	-1.04$\pm$0.09	& steep \\
13:15:03.47	&	$+$24:37:07.2	&	0.0112	&	0.031$\pm$0.001	&	0.09$\pm$0.002	&	1.53$\pm$0.03	&	-0.26$\pm$0.01	&	-0.26$\pm$0.01	& flat \\
13:27:32.43	&	$+$47:39:06.2	&	0.0604	&	0.008$\pm$0.001	&	0.68$\pm$0.05	&	1.75$\pm$0.04	&	-0.37$\pm$0.01	&	-0.37$\pm$0.01	& flat \\
13:44:42.21	&	$+$55:53:13.1	&	0.0378	&	0.145$\pm$0.005	&	5.05$\pm$0.19	&	3.74$\pm$0.05	&	-0.48$\pm$0.06	&	-0.54$\pm$0.07	& flat \\
13:47:33.42	&	$+$12:17:24.1	&	0.1217	&	5.398$\pm$0.001	&	2100.00$\pm$1.60	&	4.81$\pm$0.004	&	-0.49$\pm$0.01	&	-0.49$\pm$0.01	& flat \\
14:06:49.77	&	$+$06:10:35.1	&	0.1132	&	0.016$\pm$0.001	&	5.32$\pm$0.40	&	3.38$_{-0.20}^{+0.10}$	&	-0.69$\pm$0.22	&	-0.85$\pm$0.24	& steep \\
14:08:18.97	&	$+$19:46:22.1	&	0.1237	&	0.008$\pm$0.002	&	3.06$\pm$0.65	&	2.31$_{-0.10}^{+0.09}$	&	-0.61$\pm$0.01	&	-0.61$\pm$0.01	& steep \\
15:01:02.00	&	$+$14:20:01.7	&	0.1477	&	0.011$\pm$0.001	&	6.19$\pm$0.28	&	2.83$\pm$0.03	&	-0.15$\pm$0.01	&	-0.15$\pm$0.01	& flat \\
15:09:16.48	&	$-$11:19:25.2	&	0.0062	&	0.034$\pm$0.002	&	0.03$\pm$0.002	&	2.54$\pm$0.03	&	-1.01$\pm$0.01	&	-1.01$\pm$0.01	& steep \\
15:13:13.10	&	$+$07:13:31.2	&	0.0130	&	0.054$\pm$0.001	&	0.22$\pm$0.003	&	3.05$_{-0.10}^{+0.09}$	&	-0.04$\pm$0.20	&	-0.74$\pm$0.22	& flat \\
15:25:49.56	&	$+$05:22:49.5	&	0.0540	&	0.013$\pm$0.001	&	0.92$\pm$0.06	&	3.28$\pm$0.03	&	-0.83$\pm$0.01	&	-0.83$\pm$0.01	& steep \\
15:26:59.18	&	$+$35:58:39.2	&	0.0552	&	0.015$\pm$0.001	&	1.10$\pm$0.04	&	1.45$_{-0.03}^{+0.02}$	&	-0.20$\pm$0.01	&	-0.20$\pm$0.01	& flat \\
15:27:27.23	&	$-$09:55:50.4	&	0.0400	&	0.018$\pm$0.001	&	0.72$\pm$0.04	&	2.74$_{-0.05}^{+0.04}$	&	-0.75$\pm$0.03	&	-0.75$\pm$0.03	& steep \\
15:34:57.26	&	$+$23:30:11.1	&	0.0181	&	0.327$\pm$0.004	&	2.54$\pm$0.03	&	3.46$\pm$0.03	&	-0.30$\pm$0.04	&	-0.58$\pm$0.04	& steep \\
16:11:40.34	&	$-$01:47:06.9	&	0.1336	&	0.021$\pm$0.001	&	10.30$\pm$0.25	&	2.26$\pm$0.03	&	-0.65$\pm$0.01	&	-0.32$\pm$0.03	& steep \\
16:16:11.20	&	$+$42:24:01.4	&	0.0232	&	0.010$\pm$0.001	&	0.13$\pm$0.007	&	2.11$_{-0.20}^{+0.10}$	&	-0.49$\pm$0.25	&	-0.83$\pm$0.27	& steep \\
16:32:21.37	&	$+$15:51:47.1	&	0.2417	&	0.008$\pm$0.001	&	12.40$\pm$1.90	&	3.60$_{-0.08}^{+0.07}$	&	-0.19$\pm$0.01	&	-0.19$\pm$0.01	& flat \\
16:42:40.07	&	$-$09:43:15.7	&	0.0270	&	0.057$\pm$0.001	&	1.01$\pm$0.04	&	2.83$\pm$0.03	&	-0.74$\pm$0.01	&	-0.74$\pm$0.01	& steep \\
17:23:21.99	&	$-$00:17:02.1	&	0.0428	&	0.082$\pm$0.001	&	3.57$\pm$0.07	&	2.88$_{-0.30}^{+0.20}$	&	 0.22$\pm$0.45	&	-0.70$\pm$0.49	& peaked \\
17:54:29.69	&	$+$32:53:12.6	&	0.0260	&	0.046$\pm$0.002	&	0.75$\pm$0.03	&	2.84$_{-0.08}^{+0.06}$	&	-0.68$\pm$0.12	&	-0.89$\pm$0.14  & steep \\
18:38:35.44	&	$+$35:52:20.0	&	0.1162	&	0.021$\pm$0.001	&	6.99$\pm$0.19	&	2.48$\pm$0.05	&	 0.03$\pm$0.07	& 	 0.75$\pm$0.08	& rising \\
19:00:41.18	&	$+$35:21:25.6	&	0.1067	&	0.006$\pm$0.001	&	1.76$\pm$0.19	&	2.96$\pm$0.10	&	-0.63$\pm$0.12	&	-1.08$\pm$0.13	& steep \\
20:51:25.88	&	$+$18:58:05.0	&	0.0291	&	0.024$\pm$0.001	&	0.50$\pm$0.02	&	3.93$_{-0.06}^{+0.05}$	&	-0.79$\pm$0.08	&	-0.92$\pm$0.08	& steep \\
20:57:24.12	&	$+$17:07:41.3	&	0.0361	&	0.044$\pm$0.002	&	1.40$\pm$0.06	&	3.60$\pm$0.03	&	-0.82$\pm$0.02	&	-0.82$\pm$0.02	& steep \\
22:04:35.99	&	$+$42:19:40.1	&	0.0143	&	0.036$\pm$0.002	&	0.17$\pm$0.01	&	2.27$_{-0.05}^{+0.04}$	&	-0.45$\pm$0.06	&	-0.60$\pm$0.07	& steep \\
22:07:49.57	&	$+$30:39:42.1	&	0.1269	&	0.006$\pm$0.008	&	2.79$\pm$3.60	&	1.97$_{-0.60}^{+0.40}$	&	-0.72$\pm$0.01	&	-0.72$\pm$0.01	& steep \\
22:11:34.12	&	$-$18:17:04.3	&	0.1702	&	0.007$\pm$0.001	&	4.50$\pm$0.35	&	2.02$\pm$0.04	&	 0.34$\pm$0.01	&	 0.34$\pm$0.01	& rising \\
22:14:10.29	&	$+$04:52:26.1	&	0.1938	&	0.008$\pm$0.001	&	8.91$\pm$0.54	&	3.23$_{-0.08}^{+0.07}$	&	-0.59$\pm$0.01	&	-0.59$\pm$0.01	& steep \\
22:51:49.23	&	$-$17:52:25.4	&	0.0778	&	0.006$\pm$0.001	&	0.90$\pm$0.08	&	1.30$_{-0.05}^{+0.04}$	&	-0.40$\pm$0.01	&	-0.40$\pm$0.01	& flat \\
23:04:21.17	&	$+$34:21:49.1	&	0.1080	&	0.008$\pm$0.001	&	2.45$\pm$0.16	&	2.20$\pm$0.04	&	-0.98$\pm$0.01	&	-0.98$\pm$0.01	& steep \\
23:05:20.05	&	$+$07:41:45.8	&	0.1494	&	0.020$\pm$0.001	&	12.40$\pm$0.81	&	2.60$_{-0.10}^{+0.08}$	&	-0.93$\pm$0.15	&	-1.46$\pm$0.17	& ultra-steep \\
23:07:35.72	&	$+$04:15:59.9	&	0.0474	&	0.016$\pm$0.001	&	0.89$\pm$0.03	&	1.66$\pm$0.02	&	-0.40$\pm$0.01	&	-0.40$\pm$0.01	& flat \\
23:16:00.73	&	$+$25:33:24.2	&	0.0274	&	0.036$\pm$0.002	&	0.64$\pm$0.03	&	3.68$_{-0.08}^{+0.07}$	&	-0.63$\pm$0.05	&	-0.81$\pm$0.06	& steep \\
23:25:56.09	&	$+$10:02:48.5	&	0.1280	&	0.012$\pm$0.001	&	5.09$\pm$0.44	& 	3.19$_{-0.07}^{+0.06}$    &	-0.62$\pm$0.01	&    0.07$\pm$0.02	& upturn \\
23:35:11.87	&	$+$29:30:00.5	&	0.1070	&	0.008$\pm$0.001	&	2.49$\pm$0.15	&	2.73$\pm$0.04	&	-0.48$\pm$0.01	&	-0.48$\pm$0.01	& flat \\
23:39:01.23	&	$+$36:21:09.3	&	0.0645	&	0.029$\pm$0.001	&	3.02$\pm$0.14	&	3.59$_{-0.10}^{+0.09}$	&	-0.73$\pm$0.12	&	-0.74$\pm$0.13	& steep \\
\hline
\end{longtable}
\end{small}
\begin{longtable}{|l|l|c|c|c|c|c|c|c|}
\caption{\label{tab:control} The control sample parameters: the right ascension R.A. and declination Dec, redshift z, flux density at 1.4 GHz $S_{1.4}$, radio luminosity $P_{1.4}$ and radio loudness $\log~R$, spectral indices at 1.4 GHz $\alpha_{1.4}$ and at 4.7 GHz $\alpha_{4.7}$, type of radio spectrum.}\\
\hline
R.A.	&	Dec.	&	$z$	&	$S_{1.4}$, Jy	&	$P_{1.4}, 10^{23}$ W Hz$^{-1}$ &	$\log~R$	&			$\alpha_{1.4}$	&	$\alpha_{4.7}$	&	type	\\
\hline
1 & 2 & 3 & 4 & 5 & 6 & 7 & 8 & 9  \\
\hline
\endfirsthead
\caption[]{continued.}\\
\hline
R.A.	&	Dec.	&	$z$	&	$S_{1.4}$, Jy	&	$P_{1.4}, 10^{23}$ W Hz$^{-1}$	&	$\log~R$	&			$\alpha_{1.4}$	&	$\alpha_{4.7}$	&	type	\\
\hline
1 & 2 & 3 & 4 & 5 & 6 & 7 & 8 & 9 \\
\hline
\endhead
\hline
\endfoot
00:04:38.66	&	+36:53:30.7	&	0.1179	&	0.015$\pm$0.001	&	5.76$\pm$0.70	&	1.87$_{-0.59}^{+0.24}$	&	-0.75$\pm$0.35	&	-1.29$\pm$0.4	& steep \\
00:07:39.95	&	+27:13:38.9	&	0.1251	&	0.005$\pm$0.001	&	2.18$\pm$0.20	&	1.45$\pm$0.05	&	-0.31$\pm$0.02	&	-0.31$\pm$0.02	& flat \\
00:26:51.97	&	+34:01:22.3	&	0.1737	&	0.020$\pm$0.001	&	17.60$\pm$0.45	&	2.49$\pm$0.02	&	-0.88$\pm$0.01	&	-0.88$\pm$0.01	& steep \\
00:29:30.60	&	+24:30:08.1	&	0.1094	&	0.007$\pm$0.010	&	2.00$\pm$3.40	&	1.54$_{-0.74}^{+0.44}$	&	-0.25$\pm$0.02	&	-0.25$\pm$0.02	& flat \\
00:35:49.36	&	+27:12:39.8	&	0.1919	&	0.009$\pm$0.001	&	9.37$\pm$0.78	&	2.23$_{-0.05}^{+0.04}$	&	-0.93$\pm$0.01	&	-0.93$\pm$0.01	& steep \\
00:53:21.86	&	+04:42:36.6	&	0.1010	&	0.005$\pm$0.001	&	1.38$\pm$0.18	&	1.69$_{-0.07}^{+0.06}$	&	-0.22$\pm$0.01	&	-0.22$\pm$0.01	& flat \\
01:26:28.25	&	+35:20:11.5	&	0.1341	&	0.007$\pm$0.001	&	3.84$\pm$0.28	&	3.28$_{-0.06}^{+0.05}$	&	-1.72$\pm$0.02	&	-1.72$\pm$0.02	& ultra-steep \\
01:37:41.27	&	+33:09:35.6	&	0.3670	&	16.018$\pm$0.001&	73500.00$\pm$110.00	&	4.17$\pm$0.005	&	-0.79$\pm$0.01	&	-0.91$\pm$0.01	& steep \\
01:53:28.28	&	+26:09:39.8	&	0.3264	&	0.040$\pm$0.002	&	137.00$\pm$6.30	&	2.16$\pm$0.04	&	-0.67$\pm$0.05	&	-0.76$\pm$0.05	& steep \\
01:59:50.25	&	+00:23:38.9	&	0.1631	&	0.027$\pm$0.001	&	20.20$\pm$0.47	&	1.18$_{-0.02}^{+0.01}$	&	-0.82$\pm$0.01	&	-0.37$\pm$0.01	& steep \\
02:08:07.21	&	+08:50:02.8	&	0.3450	&	0.011$\pm$0.001	&	46.00$\pm$5.30	&	2.11$_{-0.06}^{+0.05}$	&	-1.11$\pm$0.01	&	-1.11$\pm$0.01	& steep \\
02:10:08.45	&	+23:50:50.0	&	0.1104	&	0.011$\pm$0.001	&	3.33$\pm$0.16	&	1.60$\pm$0.03	&	-0.43$\pm$0.01	&	-0.43$\pm$0.01	& flat \\
02:10:35.25	&	+23:09:16.4	&	0.0471	&	0.006$\pm$0.001	&	0.32$\pm$0.05	&	1.58$_{-0.09}^{+0.07}$	&	-0.37$\pm$0.01	&	-0.37$\pm$0.01	& flat \\
02:15:21.22	&	+26:04:19.3	&	0.1219	&	0.007$\pm$0.001	&	2.35$\pm$0.49	&	1.83$_{-0.35}^{+0.19}$	&	0.67$\pm$0.36	&	-0.22$\pm$0.39	& peaked \\
02:21:10.42	&	+23:08:25.3	&	0.1512	&	0.007$\pm$0.001	&	4.67$\pm$0.33	&	1.74$\pm$0.04	&	-0.88$\pm$0.01	&	-0.88$\pm$0.01	& steep \\
02:38:15.69	&	+19:39:49.5	&	0.1404	&	0.006$\pm$0.001	&	3.49$\pm$0.40	&	2.88$_{-0.12}^{+0.09}$	&	-0.79$\pm$0.11	&	-0.97$\pm$0.12	& steep \\
02:43:46.04	&	+04:06:36.1	&	0.1436	&	0.007$\pm$0.001	&	3.84$\pm$0.28	&	1.41$\pm$0.04	&	-0.58$\pm$0.01	&	-0.58$\pm$0.01	& steep \\
02:50:39.99	&	+27:06:52.2	&	0.1151	&	0.006$\pm$0.001	&	2.23$\pm$0.33	&	2.60$_{-0.08}^{+0.07}$	&	-0.87$\pm$0.01	&	-0.87$\pm$0.01	& steep \\
03:27:49.85	&	+16:16:58.1	&	0.1290	&	0.010$\pm$0.001	&	4.41$\pm$0.32	&	1.96$_{-0.05}^{+0.04}$	&	-0.70$\pm$0.01	&	-0.70$\pm$0.01	& steep \\
03:56:22.52	&	+26:14:53.0	&	0.1883	&	0.007$\pm$0.001	&	7.37$\pm$0.56	&	2.26$\pm$0.04	&	-1.16$\pm$0.01	&	-1.16$\pm$0.01	& ultra-steep \\
04:16:34.26	&	+12:24:57.8	&	0.2035	&	0.011$\pm$0.001	&	14.90$\pm$0.66	&	3.22$\pm$0.06	&	-1.13$\pm$0.01	&	-0.29$\pm$0.03	& steep \\
04:25:30.82	&	+01:03:05.4	&	0.1530	&	0.007$\pm$0.001	&	4.02$\pm$0.44	&	1.69$_{-0.06}^{+0.05}$	&	-0.37$\pm$0.01	&	-0.37$\pm$0.01	& flat \\
04:44:30.89	&	+26:14:10.1	&	0.1712	&	0.017$\pm$0.001	&	14.10$\pm$1.00	&	2.61$_{-0.11}^{+0.09}$	&	-0.90$\pm$0.15	&	-1.28$\pm$0.17	& ultra-steep \\
04:50:36.91	&	+06:21:39.7	&	0.1162	&	0.007$\pm$0.001	&	2.42$\pm$0.26	&	1.88$_{-0.06}^{+0.05}$	&	-0.85$\pm$0.01	&	-0.85$\pm$0.01	& steep \\
06:30:13.16	&	+35:07:55.7	&	0.1698	&	0.005$\pm$0.001	&	4.73$\pm$0.50	&	3.35$_{-0.16}^{+0.12}$	&	-1.30$\pm$0.17	&	-1.89$\pm$0.18	& ultra-steep \\
06:39:56.92	&	+28:09:56.2	&	0.1249	&	0.012$\pm$0.003	&	5.01$\pm$1.20	&	2.22$_{-0.13}^{+0.10}$	&	-0.57$\pm$0.01	&	-0.57$\pm$0.01	& steep \\
07:21:28.11	&	+04:01:46.5	&	0.1035	&	0.018$\pm$0.001	&	4.98$\pm$0.28	&	3.07$_{-0.04}^{+0.03}$	&	-0.43$\pm$0.01	&	-0.43$\pm$0.01	& flat \\
07:35:29.29	&	+04:50:27.7	&	0.1303	&	0.006$\pm$0.001	&	2.85$\pm$0.22	&	2.89$\pm$0.05	&	-0.44$\pm$0.01	&	-0.44$\pm$0.01	& flat \\
08:03:02.97	&	+07:25:39.3	&	0.1179	&	0.006$\pm$0.001	&	2.09$\pm$0.18	&	1.23$\pm$0.04	&	-0.12$\pm$0.01	&	-0.12$\pm$0.01	& flat \\
08:03:22.00	&	+07:03:33.1	&	0.1407	&	0.012$\pm$0.001	&	6.18$\pm$0.27	&	1.54$\pm$0.02	&	-0.50$\pm$0.01	&	-0.50$\pm$0.01	& flat \\
08:14:53.52	&	+04:56:14.2	&	0.1033	&	0.008$\pm$0.001	&	2.31$\pm$0.15	&	1.28$\pm$0.04	&	-0.98$\pm$0.02	&	-0.98$\pm$0.02	& steep \\
08:17:55.27	&	+31:28:27.9	&	0.1234	&	0.073$\pm$0.003	&	29.30$\pm$1.10	&	5.70$_{-0.85}^{+0.36}$	&	-0.51$\pm$0.01	&	-0.51$\pm$0.01	& steep \\
08:23:42.80	&	+31:01:13.3	&	0.1501	&	0.008$\pm$0.001	&	4.66$\pm$0.39	&	2.39$_{-0.22}^{+0.14}$	&	-0.31$\pm$0.32	&	0.27$\pm$0.35	& upturn \\
08:54:48.87	&	+20:06:30.7	&	0.3056	&	1.512$\pm$0.001	&	3400.00$\pm$5.10	&	2.92$\pm$0.01	&	0.32$\pm$0.01	&	0.32$\pm$0.01	& rising \\
08:58:41.63	&	+10:41:23.8	&	0.1480	&	0.013$\pm$0.001	&	7.87$\pm$0.34	&	1.58$_{-0.05}^{+0.04}$	&	-0.89$\pm$0.06	&	-1.31$\pm$0.06	& steep \\
09:07:34.76	&	+01:25:02.5	&	0.1019	&	0.013$\pm$0.001	&	3.59$\pm$0.15	&	1.90$\pm$0.05	&	-0.65$\pm$0.07	&	-1.13$\pm$0.08	& steep \\
09:14:13.77	&	+03:22:00.0	&	0.1451	&	0.011$\pm$0.001	&	6.22$\pm$0.29	&	1.43$_{-0.03}^{+0.02}$	&	-0.53$\pm$0.01	&	-0.53$\pm$0.01	& steep \\
09:45:21.31	&	+17:37:54.0	&	0.1282	&	0.046$\pm$0.001	&	20.90$\pm$0.32	&	2.18$_{-0.05}^{+0.04}$	&	-0.86$\pm$0.09	&	-0.96$\pm$0.09	& steep \\
09:56:58.98	&	+35:07:08.7	&	0.1004	&	0.008$\pm$0.001	&	2.23$\pm$0.14	&	1.74$\pm$0.03	&	-1.25$\pm$0.01	&	-1.25$\pm$0.01	& ultra-steep \\
10:00:24.02	&	+18:44:29.0	&	0.1076	&	0.008$\pm$0.001	&	2.70$\pm$0.32	&	1.22$_{-0.07}^{+0.06}$	&	-1.15$\pm$0.02	&	-1.15	$\pm$0.02	& ultra-steep \\
10:11:28.32	&	+26:06:52.0	&	0.1164	&	0.006$\pm$0.001	&	2.06$\pm$0.22	&	3.38$_{-0.08}^{+0.07}$	&	-0.85$\pm$0.01	&	-0.85$\pm$0.01	& steep \\
10:14:47.95	&	+16:38:48.6	&	0.1247	&	0.010$\pm$0.001	&	4.24$\pm$0.21	&	1.84$\pm$0.03	&	-0.80$\pm$0.01	&	-0.80$\pm$0.01	& steep \\
10:18:24.57	&	+15:36:38.1	&	0.1110	&	0.007$\pm$0.002	&	2.47$\pm$0.55	&	2.73$_{-0.18}^{+0.13}$	&	-1.02$\pm$0.07	&	-1.02$\pm$0.07	& steep \\
10:24:01.27	&	+00:00:41.5	&	0.1265	&	0.007$\pm$0.001	&	2.76$\pm$0.24	&	1.27$_{-0.05}^{+0.04}$	&	0.19$\pm$0.02	&	0.19$\pm$0.02	& rising \\
10:50:57.03	&	+18:53:23.1	&	0.2187	&	0.009$\pm$0.001	&	14.40$\pm$2.80	&	2.20$_{-0.46}^{+0.22}$	&	-1.34$\pm$0.49	&	-2.19$\pm$0.54	& ultra-steep \\
11:14:38.88	&	+32:41:33.4	&	0.1890	&	0.110$\pm$0.001	&	123.00$\pm$0.65	&	3.13$\pm$0.01	&	-1.17$\pm$0.01	&	-1.17$\pm$0.01	& ultra-steep \\
11:20:06.32	&	+09:01:33.0	&	0.1294	&	0.006$\pm$0.001	&	2.18$\pm$0.20	&	1.58$_{-0.06}^{+0.05}$	&	0.39$\pm$0.03	&	0.39$\pm$0.03	& rising \\
11:21:29.21	&	+11:22:23.1	&	0.1848	&	0.008$\pm$0.001	&	7.68$\pm$0.93	&	1.46$_{-0.07}^{+0.06}$	&	-0.99$\pm$0.02	&	-0.99$\pm$0.02	& steep \\
11:26:02.48	&	+34:34:49.7	&	0.1108	&	0.006$\pm$0.001	&	1.83$\pm$0.17	&	1.43$_{-0.05}^{+0.04}$	&	-0.90$\pm$0.01	&	-0.90$\pm$0.01	& steep \\
11:44:10.67	&	+09:10:42.5	&	0.1068	&	0.008$\pm$0.001	&	2.53$\pm$0.15	&	3.78$_{-0.10}^{+0.08}$	&	-0.70$\pm$0.01	&	-0.70$\pm$0.01	& steep \\
11:53:14.22	&	+13:14:27.9	&	0.1273	&	0.014$\pm$0.002	&	6.20$\pm$0.90	&	1.93$_{-0.09}^{+0.08}$	&	-0.86$\pm$0.01	&	-0.86$\pm$0.01	& steep \\
12:02:05.58	&	+11:28:10.2	&	0.1937	&	0.009$\pm$0.001	&	10.80$\pm$1.20	&	1.64$_{-0.07}^{+0.06}$	&	-1.06$\pm$0.03	&	-1.06$\pm$0.03	& steep \\
12:13:20.00	&	+31:40:53.2	&	0.2065	&	0.032$\pm$0.001	&	37.90$\pm$0.99	&	2.34$\pm$0.03	&	-0.51$\pm$0.01	&	-0.51$\pm$0.01	& steep \\
12:13:39.57	&	+28:32:14.6	&	0.1031	&	0.007$\pm$0.001	&	1.89$\pm$0.13	&	1.07$_{-0.04}^{+0.03}$	&	0.10$\pm$0.01	&	0.10$\pm$0.01	& rising \\
12:29:06.41	&	+02:03:05.1	&	0.1584	&	54.992$\pm$1.90	&	34500.00$\pm$1200.00	&	3.35$\pm$0.02	&	0.03$\pm$0.01	&	-0.13$\pm$0.01	& peaked \\
12:54:00.78	&	+10:11:12.8	&	0.3190	&	0.009$\pm$0.001	&	27.10$\pm$1.60	&	1.93$\pm$0.03	&	-0.55$\pm$0.01	&	-0.55$\pm$0.01	& steep \\
12:57:39.37	&	+08:09:30.0	&	0.2699	&	0.007$\pm$0.001	&	14.80$\pm$1.20	&	1.29$\pm$0.04	&	-0.29$\pm$0.01	&	-0.29$\pm$0.01	& flat \\
13:06:00.59	&	+00:01:31.9	&	0.1371	&	0.010$\pm$0.001	&	5.36$\pm$0.27	&	4.38$_{-0.26}^{+0.16}$	&	-1.13$\pm$0.01	&	-1.13$\pm$0.01	& ultra-steep \\
13:08:51.81	&	+20:41:23.8	&	0.1142	&	0.011$\pm$0.001	&	3.59$\pm$0.17	&	1.35$\pm$0.03	&	-0.38$\pm$0.01	&	-0.38$\pm$0.01	& flat \\
13:16:53.96	&	+23:40:47.9	&	0.1380	&	0.014$\pm$0.001	&	6.72$\pm$0.30	&	2.09$\pm$0.03	&	-0.30$\pm$0.01	&	-0.30$\pm$0.01	& flat \\
13:18:10.28	&	+04:19:16.7	&	0.1133	&	0.012$\pm$0.001	&	3.96$\pm$0.18	&	1.42$\pm$0.03	&	-0.78$\pm$0.02	&	-0.78$\pm$0.02	& steep \\
13:20:37.99	&	+01:17:28.4	&	0.1034	&	0.007$\pm$0.001	&	2.04$\pm$0.30	&	1.23$_{-0.08}^{+0.07}$	&	-0.49$\pm$0.01	&	-0.49$\pm$0.01	& flat \\
13:37:18.73	&	+24:23:02.8	&	0.1076	&	0.020$\pm$0.001	&	6.03$\pm$0.16	&	0.95$\pm$0.03	&	-0.54$\pm$0.04	&	-0.83$\pm$0.04	& steep \\
13:40:15.29	&	+33:24:37.5	&	0.2473	&	0.008$\pm$0.001	&	16.30$\pm$1.60	&	2.34$\pm$0.05	&	-0.93$\pm$0.01	&	-0.93$\pm$0.01	& steep \\
13:50:16.26	&	+16:28:07.7	&	0.1124	&	0.007$\pm$0.001	&	2.46$\pm$0.17	&	1.27$\pm$0.04	&	-0.97$\pm$0.01	&	-0.97$\pm$0.01	& steep \\
13:53:31.40	&	+04:28:04.4	&	0.1360	&	0.010$\pm$0.001	&	4.91$\pm$0.24	&	1.82$\pm$0.03	&	-0.27$\pm$0.01	&	-0.27$\pm$0.01	& flat \\
13:56:09.95	&	+29:05:35.7	&	0.1085	&	0.012$\pm$0.001	&	3.73$\pm$0.16	&	1.84$_{-0.03}^{+0.02}$	&	-0.58$\pm$0.01	&	-0.58$\pm$0.01	& steep \\
13:56:46.14	&	+10:26:09.7	&	0.1230	&	0.065$\pm$0.002	&	25.40$\pm$1.20	&	1.92$_{-0.11}^{+0.09}$	&	-0.33$\pm$0.18	&	-0.96$\pm$0.20	& steep \\
14:06:38.17	&	+01:02:55.6	&	0.2365	&	0.015$\pm$0.001	&	25.10$\pm$1.20	&	2.15$\pm$0.03	&	-0.68$\pm$0.02	&	-0.68$\pm$0.02	& steep \\
14:08:19.06	&	+29:04:46.7	&	0.1168	&	0.010$\pm$0.006	&	3.55$\pm$2.00	&	1.21$_{-0.38}^{+0.20}$	&	-0.74$\pm$0.01	&	-0.74$\pm$0.01	& steep \\
14:20:56.67	&	-00:04:29.7	&	0.1020	&	0.007$\pm$0.001	&	1.82$\pm$0.13	&	1.74$\pm$0.04	&	-0.48$\pm$0.01	&	-0.48$\pm$0.01	& flat \\
14:22:31.12	&	+26:02:03.3	&	0.1588	&	0.010$\pm$0.001	&	7.33$\pm$0.37	&	1.48$\pm$0.03	&	-0.85$\pm$0.02	&	-0.85$\pm$0.02	& steep \\
14:33:27.54	&	+28:12:01.2	&	0.1749	&	0.006$\pm$0.002	&	5.16$\pm$1.30	&	1.43$_{-0.13}^{+0.10}$	&	-0.56$\pm$0.01	&	-0.56$\pm$0.01	& steep \\
14:42:44.78	&	+26:21:34.6	&	0.1074	&	0.007$\pm$0.002	&	2.02$\pm$0.56	&	1.37$_{-0.15}^{+0.11}$	&	-0.87$\pm$0.02	&	-0.87$\pm$0.02	& steep \\
14:49:19.32	&	+13:50:32.6	&	0.1039	&	0.008$\pm$0.001	&	2.12$\pm$0.14	&	1.52$_{-0.04}^{+0.03}$	&	-0.51$\pm$0.01	&	-0.51$\pm$0.01	& steep \\
14:50:54.04	&	+35:08:38.1	&	0.2057	&	0.007$\pm$0.001	&	6.16$\pm$1.10	&	1.20$_{-0.57}^{+0.38}$	&	0.86$\pm$1.31	&	-1.57$\pm$1.42	& peaked \\
14:56:08.47	&	+17:18:35.4	&	0.1035	&	0.013$\pm$0.001	&	4.23$\pm$0.29	&	1.72$_{-0.04}^{+0.03}$	&	-1.9$\pm$0.01	&	-1.90$\pm$0.01	& ultra-steep \\
14:57:31.01	&	+07:03:33.6	&	0.1401	&	0.006$\pm$0.001	&	3.04$\pm$0.32	&	1.28$\pm$0.05	&	-0.52$\pm$0.01	&	-0.52$\pm$0.01	& steep \\
14:59:36.55	&	+32:44:59.3	&	0.1138	&	0.007$\pm$0.001	&	1.65$\pm$0.16	&	1.03$_{-0.12}^{+0.10}$	&	2.54$\pm$0.16	&	-0.13$\pm$0.18	& rising \\
15:02:31.79	&	+14:21:34.7	&	0.1627	&	0.017$\pm$0.001	&	12.40$\pm$0.75	&	1.81$_{-0.04}^{+0.03}$	&	-0.63$\pm$0.02	&	-0.68$\pm$0.02	& steep \\
15:09:13.86	&	+17:57:12.1	&	0.1707	&	0.012$\pm$0.001	&	10.00$\pm$0.67	&	2.09$_{-0.04}^{+0.03}$	&	-0.75$\pm$0.01	&	-0.75$\pm$0.01	& steep \\
15:17:58.50	&	+27:36:14.8	&	0.1601	&	0.007$\pm$0.001	&	4.65$\pm$0.32	&	1.91$\pm$0.04	&	0.17$\pm$0.02	&	0.17$\pm$0.02	& rising \\
15:19:23.74	&	+00:35:00.9	&	0.1539	&	0.015$\pm$0.001	&	9.88$\pm$0.34	&	1.97$\pm$0.02	&	-0.91$\pm$0.01	&	-0.91$\pm$0.01	& steep \\
15:22:37.80	&	+33:31:34.6	&	0.1244	&	0.011$\pm$0.001	&	4.70$\pm$0.26	&	1.23$_{-0.14}^{+0.11}$	&	-0.75$\pm$0.23	&	-1.57$\pm$0.26	& ultra-steep \\
15:24:43.69	&	+23:40:11.6	&	0.1390	&	0.007$\pm$0.001	&	3.39$\pm$0.50	&	1.44$_{-0.25}^{+0.16}$	&	-0.14$\pm$0.31	&	-0.70$\pm$0.34	& steep \\
15:46:21.31	&	+04:29:08.4	&	0.2370	&	0.005$\pm$0.001	&	9.02$\pm$0.84	&	2.18$\pm$0.05	&	-0.67$\pm$0.01	&	-0.67$\pm$0.01	& steep \\
16:06:37.05	&	+27:24:57.1	&	0.1139	&	0.010$\pm$0.001	&	3.45$\pm$0.24	&	1.15$_{-0.24}^{+0.15}$	&	-0.96$\pm$0.37	&	-1.14$\pm$0.40	& steep \\
16:14:13.12	&	+26:04:15.2	&	0.1310	&	0.018$\pm$0.001	&	8.57$\pm$0.26	&	1.18$_{-0.04}^{+0.03}$	&	-0.80$\pm$0.05	&	-1.05	$\pm$0.06 & steep \\
16:16:44.06	&	+03:14:19.2	&	0.1059	&	0.008$\pm$0.001	&	2.38$\pm$0.16	&	1.75$_{-0.06}^{+0.05}$	&	-1.39$\pm$0.05	&	-1.39$\pm$0.05	& ultra-steep \\
16:18:09.40	&	+01:39:22.1	&	0.1320	&	0.009$\pm$0.001	&	3.92$\pm$0.23	&	3.18$\pm$0.04	&	-0.40$\pm$0.01	&	-0.40$\pm$0.01	& flat \\
16:30:52.17	&	+04:35:54.3	&	0.1238	&	0.006$\pm$0.001	&	2.55$\pm$0.75	&	1.3$_{-0.16}^{+0.12}$	&	-0.80$\pm$0.01	&	-0.80$\pm$0.01	& steep \\
16:49:14.31	&	+34:25:11.7	&	0.1115	&	0.012$\pm$0.001	&	3.67$\pm$0.21	&	1.97$_{-0.07}^{+0.06}$	&	-0.36$\pm$0.09	&	-0.78$\pm$0.09	& steep \\
17:04:51.77	&	+02:28:41.9	&	0.1385	&	0.006$\pm$0.001	&	2.93$\pm$0.70	&	1.79$_{-0.13}^{+0.10}$	&	-0.83$\pm$0.02	&	-0.83$\pm$0.02	& steep \\
17:13:36.32	&	+20:56:04.1	&	0.1213	&	0.009$\pm$0.001 &	3.46$\pm$0.20	&	1.36$\pm$0.03	&	-0.65$\pm$0.01	&	-0.65$\pm$0.01	& steep \\
17:51:05.59	&	+26:59:03.4	&	0.1450	&	0.142$\pm$0.001	&	82.10$\pm$0.34	&	3.21$\pm$0.01	&	-0.71$\pm$0.01	&	-0.71$\pm$0.01	& steep \\
17:56:58.39	&	+13:56:08.5	&	0.1205	&	0.009$\pm$0.001	&	3.46$\pm$0.20	&	2.26$\pm$0.03	&	-0.69$\pm$0.01	&	-0.69$\pm$0.01	& steep \\
17:59:53.90	&	+06:29:10.0	&	0.1096	&	0.015$\pm$0.004	&	4.64$\pm$1.20	&	3.32$_{-0.15}^{+0.11}$	&	-0.68$\pm$0.01	&	-0.68$\pm$0.01	& steep \\
18:06:13.63	&	+21:41:32.0	&	0.1016	&	0.006$\pm$0.001	&	1.52$\pm$0.32	&	2.42$_{-0.11}^{+0.09}$	&	-0.50$\pm$0.01	&	-0.50$\pm$0.01	& flat \\
18:17:00.26	&	+15:54:50.0	&	0.1024	&	0.053$\pm$0.002	&	15.00$\pm$0.63	&	2.44$_{-0.07}^{+0.06}$	&	-0.91$\pm$0.11	&	-1.07$\pm$0.12	& steep \\
18:33:36.03	&	+22:52:01.7	&	0.1310	&	0.016$\pm$0.001	&	7.45$\pm$0.24	&	3.41$\pm$0.04	&	-0.71$\pm$0.01	&	-0.71$\pm$0.01	& steep \\
18:42:28.24	&	+36:01:15.0	&	0.1039	&	0.007$\pm$0.001	&	1.92$\pm$0.20	&	3.00$_{-0.09}^{+0.08}$	&	-0.64$\pm$0.01	&	-0.64	$\pm$0.01	& steep \\
19:48:15.89	&	+09:52:03.2	&	0.1000	&	0.015$\pm$0.001	&	3.95$\pm$0.26	&	2.44$_{-0.04}^{+0.03}$	&	-0.52$\pm$0.01	&	-0.52$\pm$0.01	& steep \\
19:58:11.01	&	+16:26:16.8	&	0.1400	&	0.051$\pm$0.002	&	27.20$\pm$1.20	&	4.64$_{-0.09}^{+0.07}$	&	-0.70$\pm$0.01	&	-0.70	$\pm$0.01	& steep \\
20:11:36.11	&	+01:37:58.5	&	0.1015	&	0.012$\pm$0.001	&	3.35$\pm$0.27	&	2.09$\pm$0.04	&	-0.66$\pm$0.01	&	-0.66$\pm$0.01	& steep \\
20:27:12.58	&	+01:16:20.7	&	0.1149	&	0.006$\pm$0.001	&	2.38$\pm$0.19	&	2.89$_{-0.08}^{+0.07}$	&	-1.19$\pm$0.01	&	-1.19$\pm$0.01	& ultra-steep \\
20:41:38.76	&	+23:12:59.3	&	0.1053	&	0.007$\pm$0.001	&	2.00$\pm$0.37	&	2.23$_{-0.10}^{+0.08}$	&	-1.06$\pm$0.01	&	-1.06	$\pm$0.01	& steep \\
20:42:00.12	&	+27:56:14.7	&	0.1025	&	0.015$\pm$0.001	&	3.91$\pm$0.14	&	1.61$\pm$0.04	&	-0.22$\pm$0.06	&	0.27$\pm$0.06	& upturn \\
20:47:30.35	&	+09:24:18.8	&	0.1218	&	0.010$\pm$0.001	&	3.79$\pm$0.39	&	1.79$_{-0.06}^{+0.05}$	&	-0.45$\pm$0.02	&	-0.45$\pm$0.02	& flat \\
20:48:17.29	&	+19:36:55.9	&	0.1807	&	0.019$\pm$0.001	&	17.90$\pm$0.49	&	3.55$_{-0.06}^{+0.05}$	&	-0.80$\pm$0.01	&	-0.80$\pm$0.01	& steep \\
20:57:19.66	&	+24:53:36.6	&	0.1730	&	0.014$\pm$0.001	&	12.60$\pm$0.74	&	2.29$_{-0.15}^{+0.11}$	&	-0.96$\pm$0.24	&	-1.20$\pm$0.25	& steep \\
21:05:05.38	&	+10:54:04.0	&	0.1072	&	0.010$\pm$0.001	&	3.39$\pm$0.36	&	1.41$\pm$0.05	&	-1.37$\pm$0.01	&	-1.37$\pm$0.01	& ultra-steep \\
21:08:43.86	&	+22:07:37.1	&	0.1076	&	0.005$\pm$0.002	&	1.68$\pm$0.68	&	4.23$_{-1.12}^{+0.49}$	&	-0.83$\pm$0.02	&	-0.83$\pm$0.02	& steep \\
21:15:59.77	&	+06:05:21.9	&	0.1058	&	0.011$\pm$0.001	&	3.21$\pm$0.34	&	2.6$_{-0.08}^{+0.07}$	&	-0.85$\pm$0.05	&	-0.85$\pm$0.05	& steep \\
21:19:10.36	&	+08:31:44.6	&	0.1015	&	0.008$\pm$0.004	&	2.11$\pm$1.00	&	4.30$_{-0.55}^{+0.24}$	&	-0.67$\pm$0.04	&	-0.67$\pm$0.04	& steep \\
21:46:33.52	&	+35:48:35.3	&	0.1518	&	0.015$\pm$0.001	&	9.05$\pm$0.35	&	3.37$_{-0.09}^{+0.07}$	&	-0.45$\pm$0.08	&	-0.72$\pm$0.08	& steep \\
21:50:16.13	&	+05:16:03.5	&	0.1710	&	0.007$\pm$0.001	&	6.34$\pm$0.47	&	1.87$\pm$0.04	&	-1.39$\pm$0.02	&	-1.39$\pm$0.02	& ultra-steep \\
22:30:46.92	&	+36:10:38.8	&	0.1175	&	0.006$\pm$0.001	&	2.28$\pm$0.18	&	1.07$\pm$0.04	&	-0.38$\pm$0.01	&	-0.38$\pm$0.01	& flat \\
22:45:09.70	&	+32:31:28.9	&	0.1572	&	0.009$\pm$0.001	&	5.89$\pm$0.34	&	1.49$\pm$0.03	&	-0.52$\pm$0.01	&	-0.52$\pm$0.01	& steep \\
22:56:42.28	&	+08:49:18.7	&	0.1661	&	0.006$\pm$0.001	&	4.20$\pm$0.46	&	1.85$_{-0.08}^{+0.06}$	&	-0.38$\pm$0.05	&	-0.38$\pm$0.05	& flat \\
23:00:49.95	&	+17:19:31.2	&	0.1191	&	0.013$\pm$0.002	&	4.82$\pm$0.58	&	1.46$_{-0.06}^{+0.05}$	&	-0.76$\pm$0.01	&	-0.76$\pm$0.01	& steep \\
23:04:20.58	&	+03:49:49.0	&	0.1185	&	0.007$\pm$0.001	&	2.71$\pm$0.23	&	3.86$_{-0.16}^{+0.12}$	&	-0.70$\pm$0.04	&	-0.70$\pm$0.04	& steep \\
23:08:34.02	&	+05:21:29.2	&	0.1730	&	0.007$\pm$0.001	&	5.49$\pm$0.89	&	1.25$_{-0.08}^{+0.07}$	&	-0.41$\pm$0.01	&	-0.41$\pm$0.01	& flat \\
23:13:54.47	&	+03:30:55.1	&	0.3053	&	0.048$\pm$0.001	&	144.00$\pm$1.90	&	3.18$\pm$0.02	&	-0.81$\pm$0.02	&	-0.81$\pm$0.02	& steep \\
23:16:35.19	&	+04:05:17.8	&	0.2201	&	4.676$\pm$0.165	&	7030.00$\pm$250.00	&	4.31$\pm$0.02	&	-0.96$\pm$0.01	&	-1.05$\pm$0.01	& steep \\
23:24:27.72	&	+29:35:41.7	&	0.2401	&	0.006$\pm$0.001	&	9.44$\pm$0.75	&	2.04$\pm$0.04	&	0.04$\pm$0.01	&	0.04$\pm$0.01	& rising \\
23:25:49.43	&	+28:34:21.1	&	0.1140	&	0.036$\pm$0.001	&	12.20$\pm$0.24	&	1.95$_{-0.04}^{+0.03}$	&	-0.65$\pm$0.06	&	-0.85$\pm$0.07	& steep \\
23:52:25.96	&	+24:40:14.1	&	0.2120	&	0.007$\pm$0.001	&	9.33$\pm$0.98	&	2.06$_{-0.06}^{+0.05}$	&	-0.92$\pm$0.02	&	-0.92$\pm$0.02	& steep \\
23:56:30.16	&	+23:38:49.2	&	0.2660	&	0.014$\pm$0.001	&	29.90$\pm$1.20	&	2.12$_{-0.05}^{+0.04}$	&	-0.60$\pm$0.06	&	-0.45$\pm$0.07	& steep \\
\hline
\end{longtable}

\clearpage
\newpage

 \begin{figure}
 \centerline{\includegraphics[width=145mm]{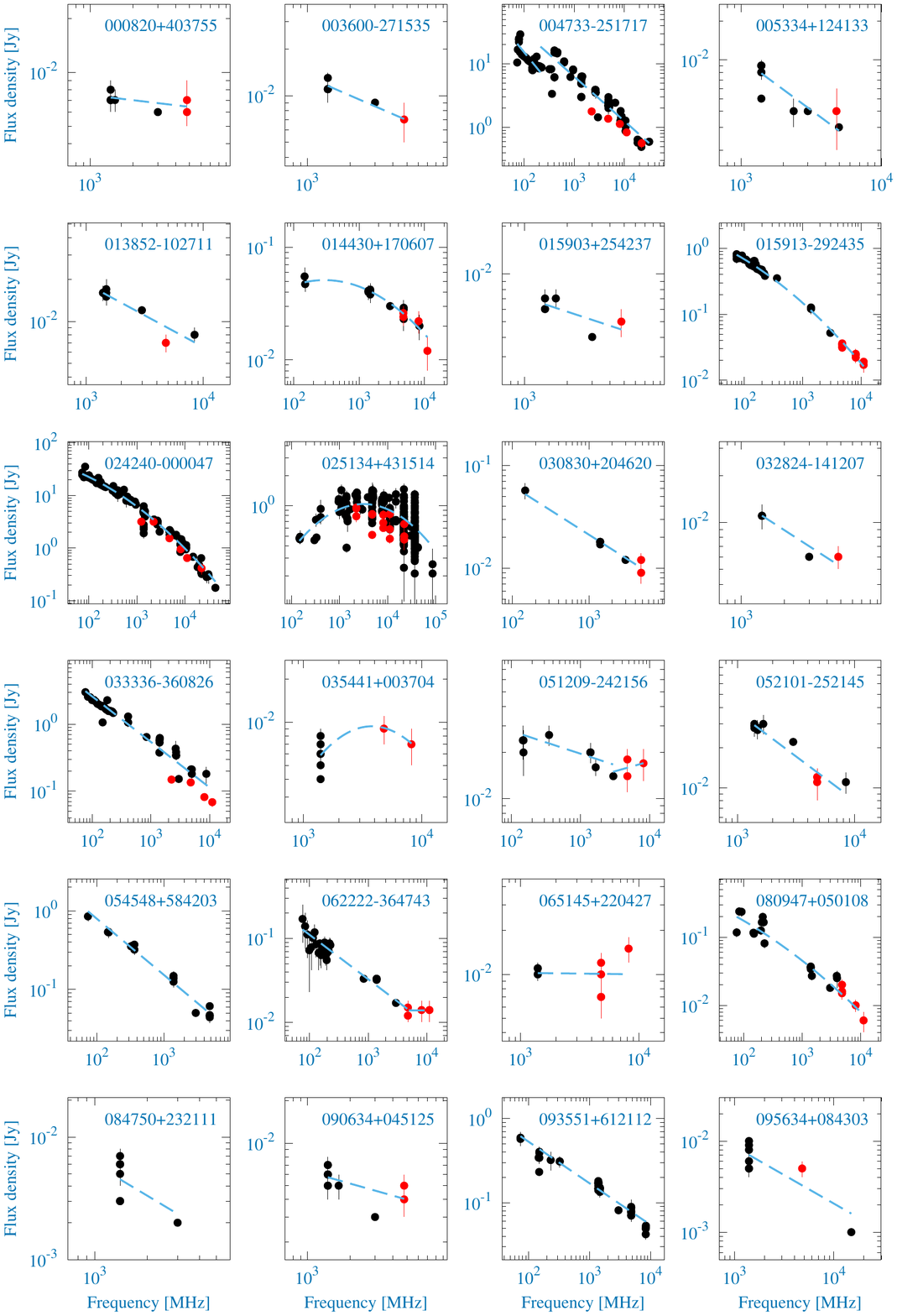}}
 \caption{Radio spectra of the OHM sample.}
 \label{fig:spectra1}
 \end{figure}
 \addtocounter{figure}{-1}
 \clearpage
 \newpage

 \begin{figure}
 \centerline{\includegraphics[width=145mm]{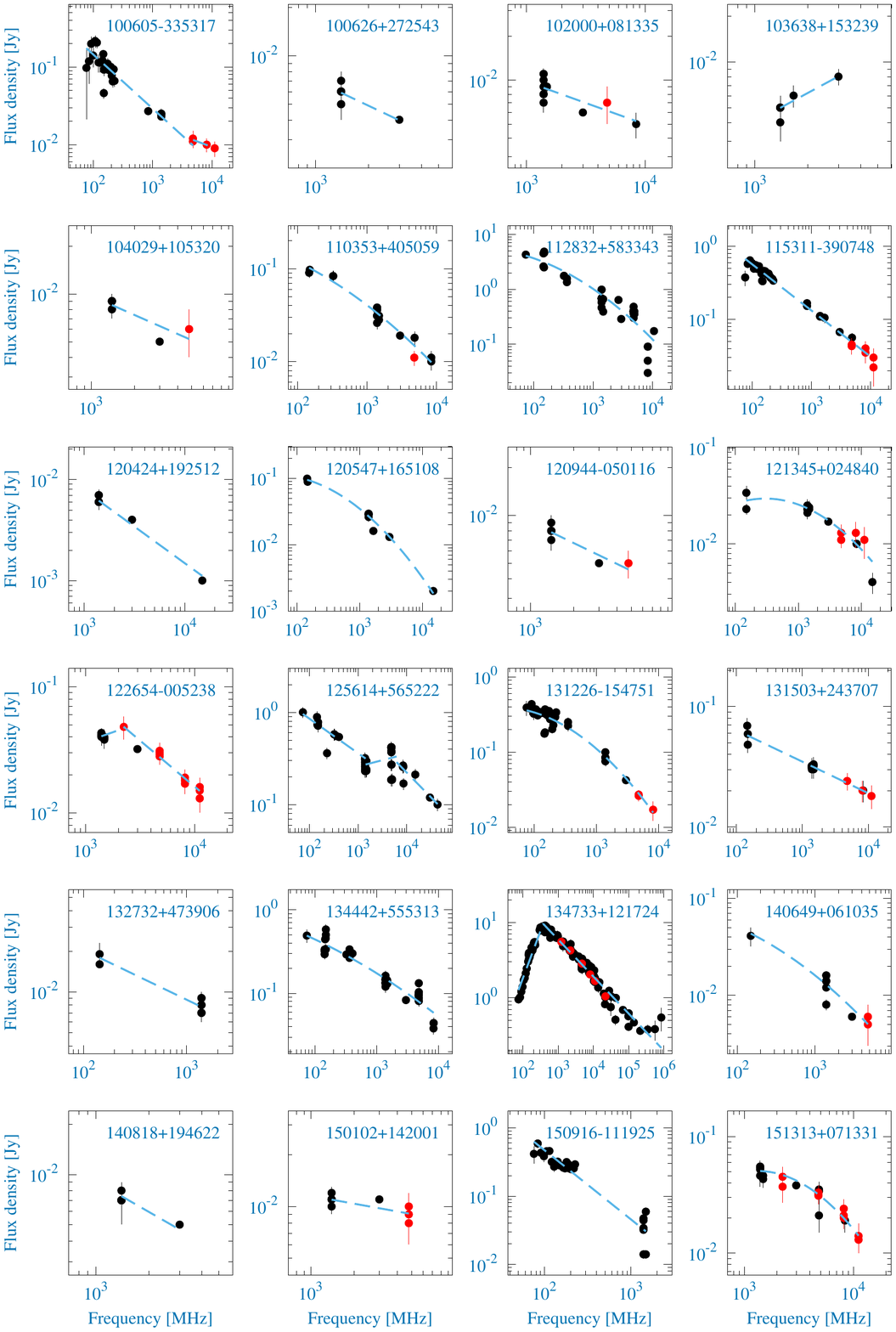}}
 \caption{Radio spectra of the OHM sample.}
 \label{fig:spectra2}
 \end{figure}
 \addtocounter{figure}{-1}
 \clearpage
 \newpage

 \begin{figure}
 \centerline{\includegraphics[width=145mm]{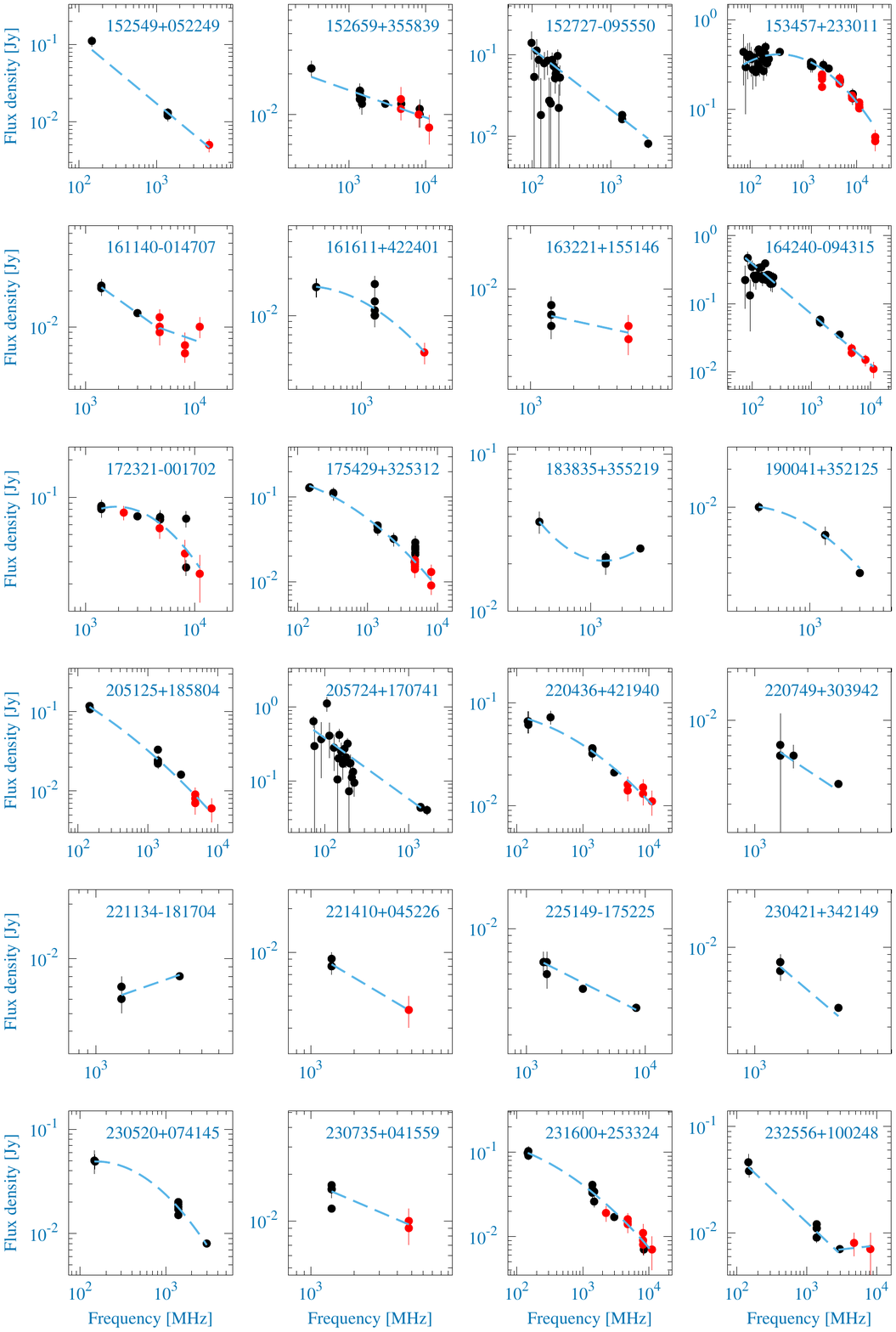}}
 \caption{Radio spectra of the OHM sample.}
 \label{fig:spectra3}
 \end{figure}
 \addtocounter{figure}{-1}
 \clearpage
 \newpage

 \begin{figure}
 \centerline{\includegraphics[width=145mm]{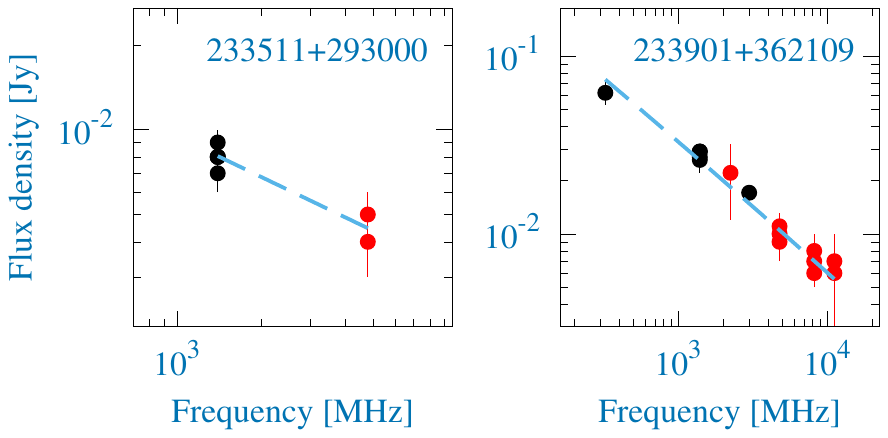}}
 \caption{Radio spectra of the OHM sample.}
 \label{fig:spectra4}
 \end{figure}


 \clearpage
 \newpage

\begin{figure}
 \centerline{\includegraphics[width=145mm]{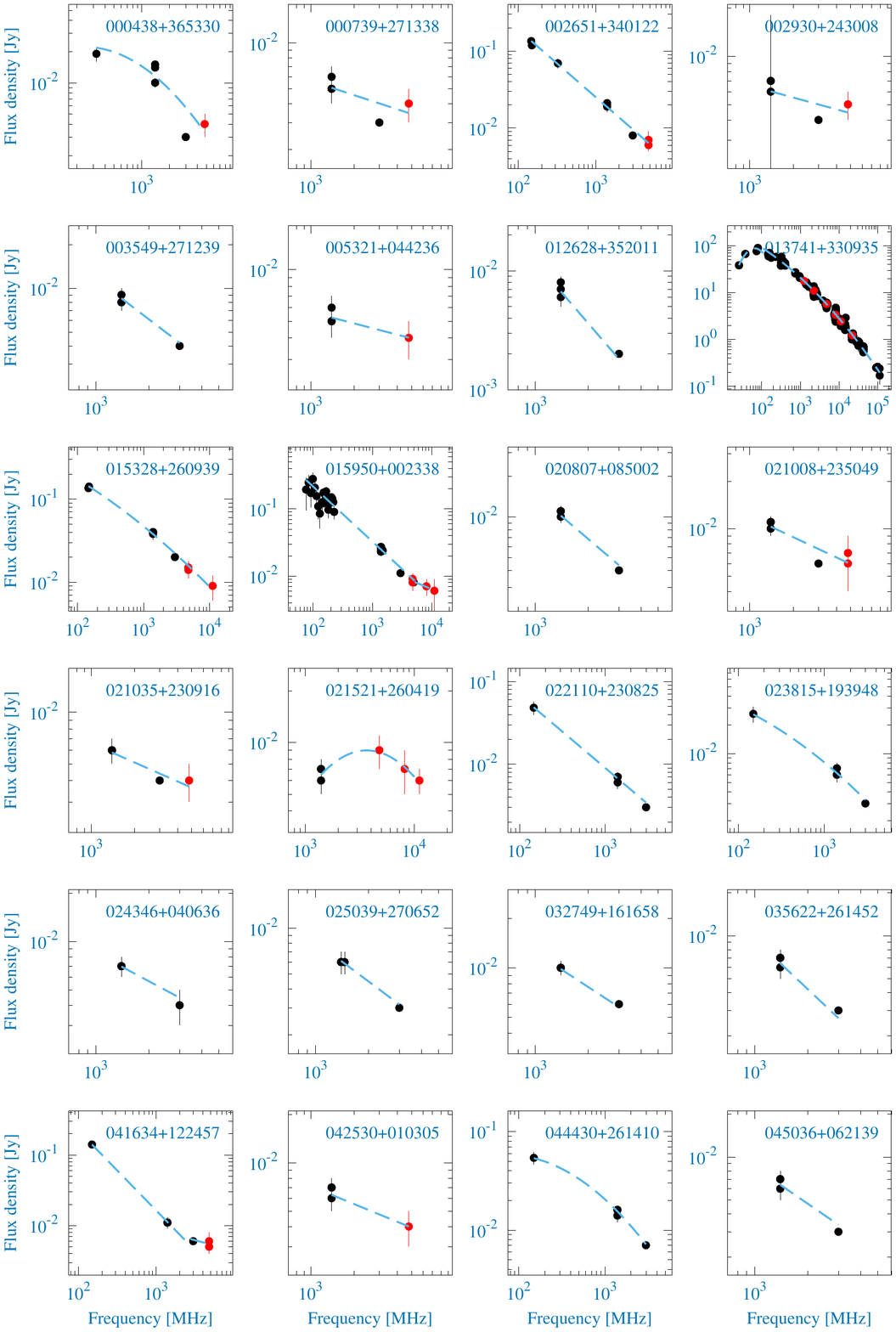}}
 \caption{Radio spectra of the control sample.}
 \label{fig:spectra5}
 \end{figure}
 \addtocounter{figure}{-1}
 \clearpage
 \newpage

 \begin{figure}
 \centerline{\includegraphics[width=145mm]{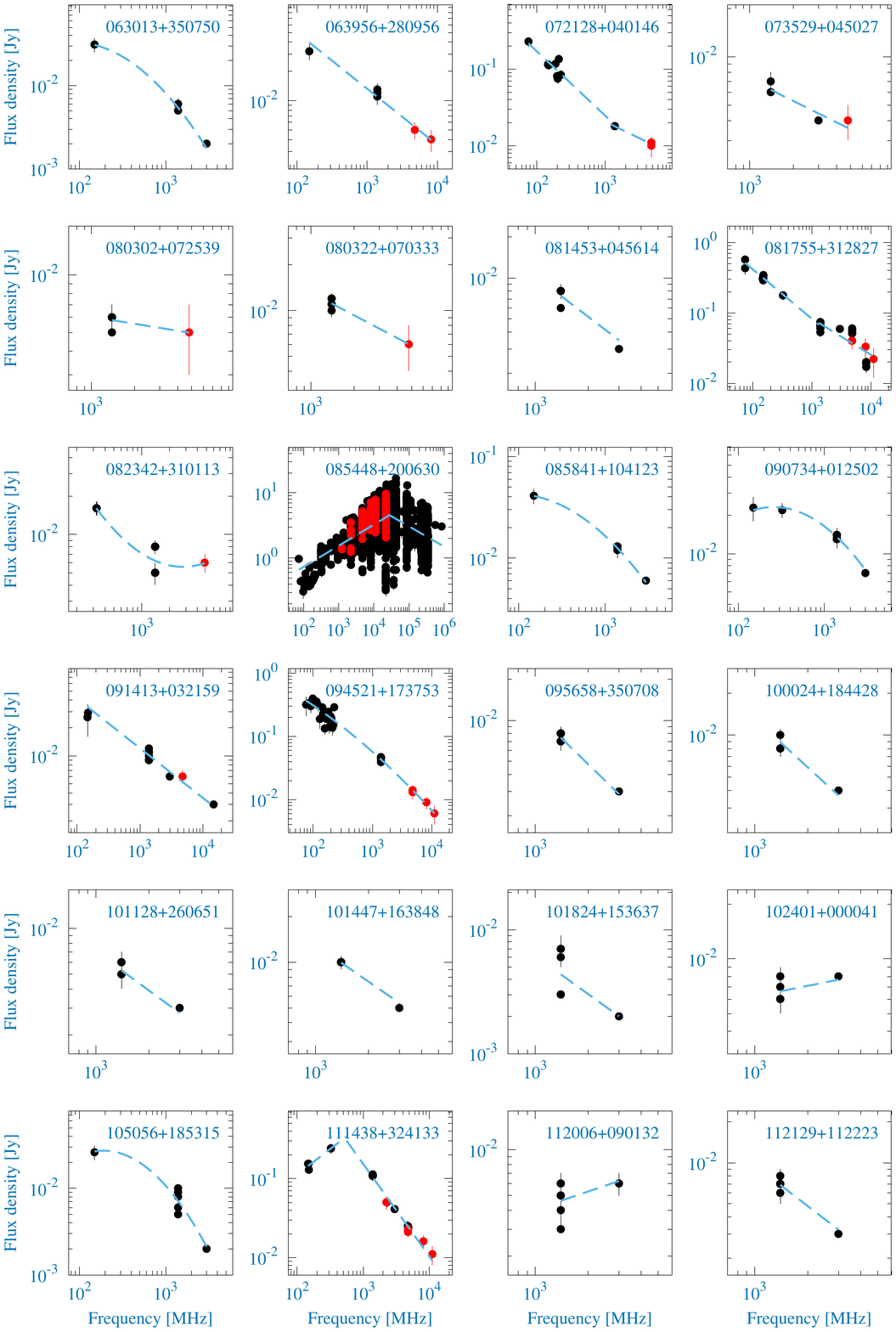}}
 \caption{Radio spectra of the control sample.}
 \label{fig:spectra6}
 \end{figure}
 \addtocounter{figure}{-1}
 \clearpage
 \newpage

 \begin{figure}
 \centerline{\includegraphics[width=145mm]{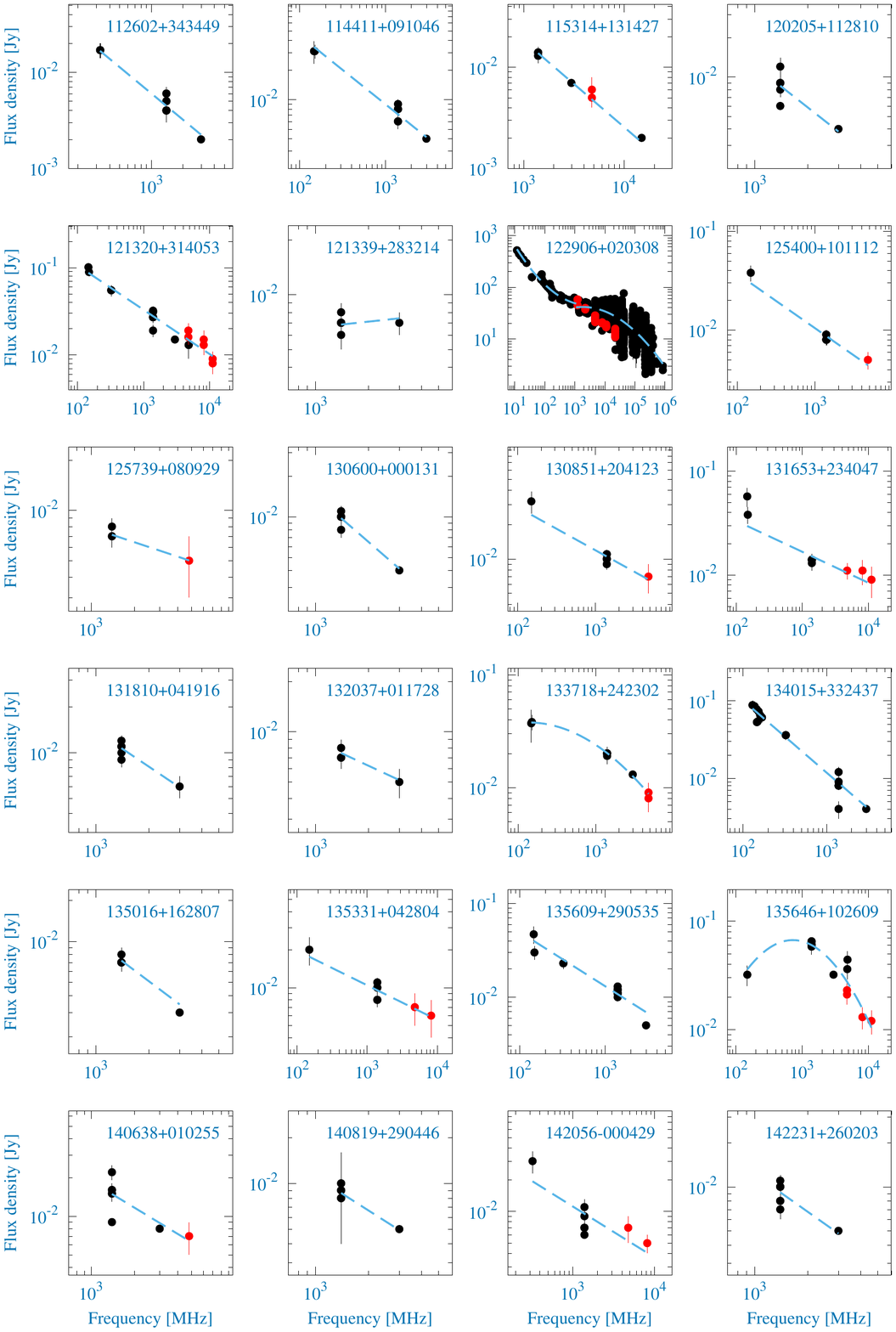}}
 \caption{Radio spectra of the control sample.}
 \label{fig:spectra7}
 \end{figure}
 \addtocounter{figure}{-1}
 \clearpage
 \newpage

\begin{figure}
 \centerline{\includegraphics[width=145mm]{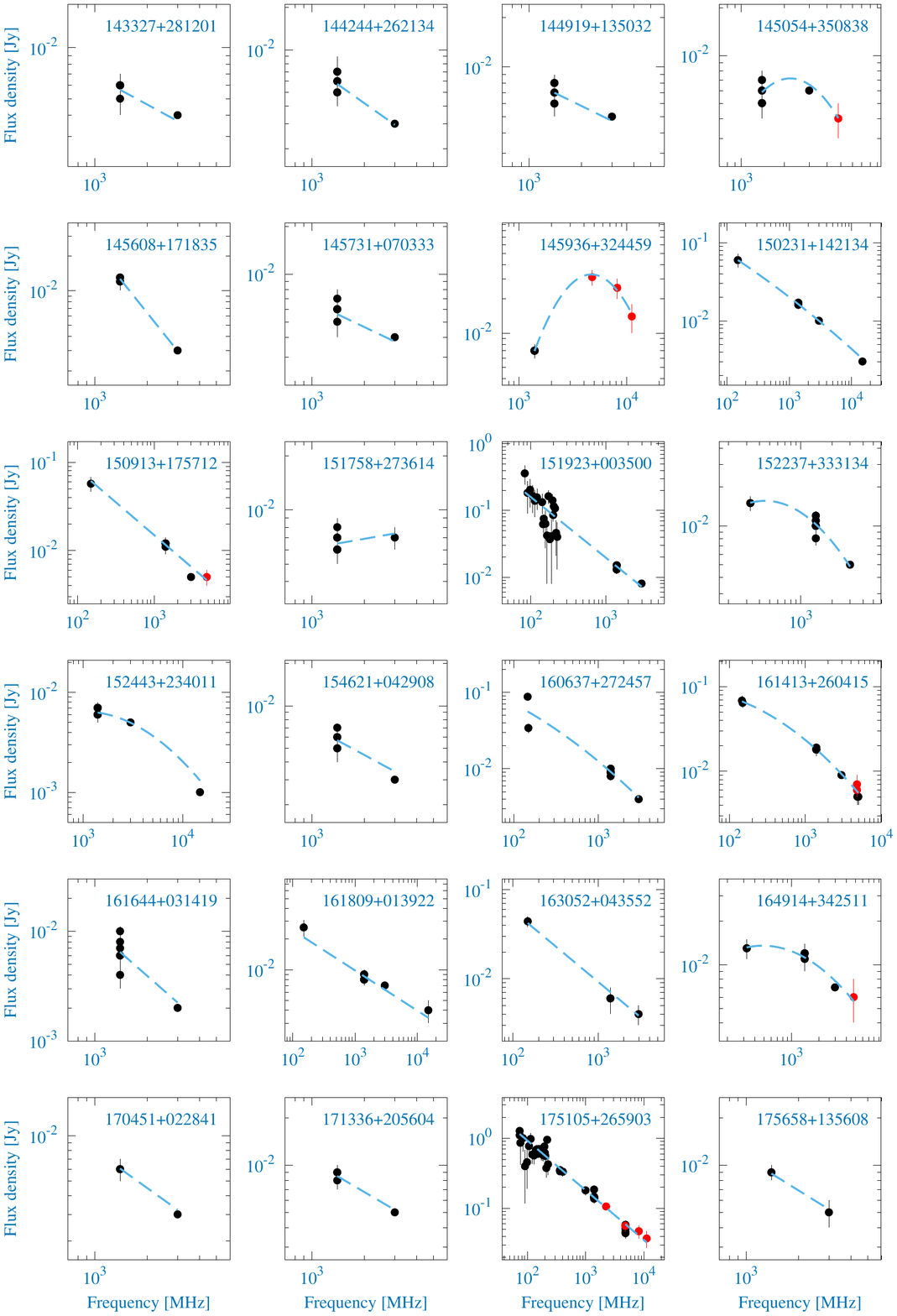}}
 \caption{Radio spectra of the control sample.}
 \label{fig:spectra8}
 \end{figure}
 \addtocounter{figure}{-1}
 \clearpage
 \newpage

 \begin{figure}
 \centerline{\includegraphics[width=143mm]{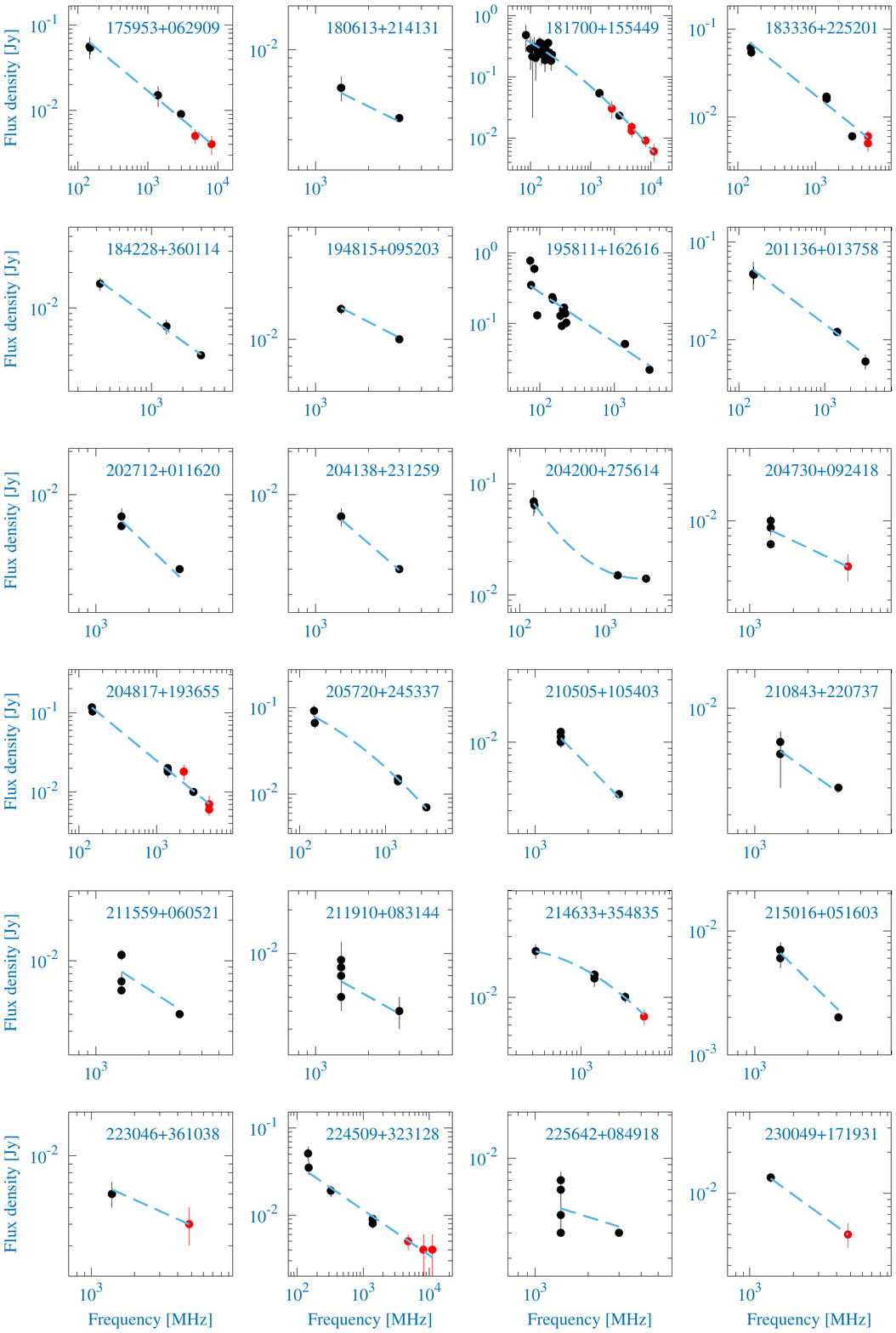}}
 \caption{Radio spectra of the control sample.}
 \label{fig:spectra9}
 \end{figure}
 \addtocounter{figure}{-1}
 \clearpage
 \newpage

\begin{figure}
 \centerline{\includegraphics[width=143mm]{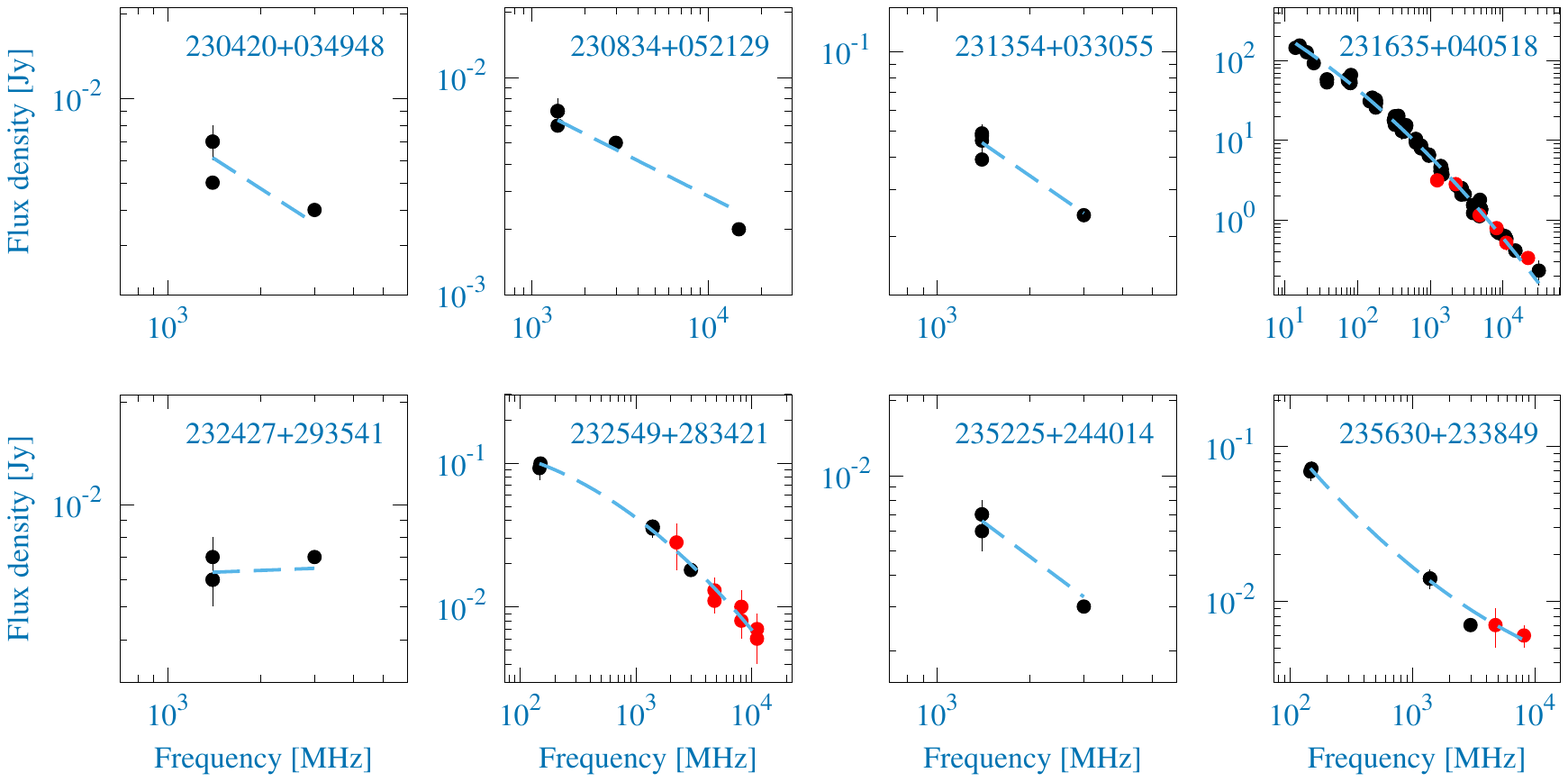}}
 \caption{Radio spectra of the control sample.}
 \label{fig:spectra10}
 \end{figure}
 \addtocounter{figure}{-1}

\bsp	
\label{lastpage}
\end{document}